%% file: top_hnl.tex
\documentclass[11pt,a4paper]{article}
\usepackage{jheppub}
\pdfoutput=1

\usepackage[utf8]{inputenc}
\usepackage{amssymb}
\usepackage{amsmath}
\usepackage{amsfonts}
\usepackage{graphicx}
\usepackage{xcolor}
\usepackage{xspace}
\usepackage[normalem]{ulem} %\usepackage[normalem]{ulem} if want italics
\usepackage{lscape}
\usepackage{ wasysym }
\usepackage{float}
\usepackage{mathrsfs}
\usepackage{slashed}
\usepackage{multirow,rotating}
\usepackage{siunitx}

\usepackage{hyperref}
\hypersetup{
  colorlinks=true,
  allcolors=darkpurple,
  pdfborder={0 0 0},
  linktocpage=true%false
}

\definecolor{darkpurple}{rgb}{0.5,0,0.5}
\definecolor{cambridgeblue}{rgb}{0.64, 0.76, 0.68}
\definecolor{darkraspberry}{rgb}{0.53, 0.15, 0.34}

\def\gsim{\raise0.3ex\hbox{$\;>$\kern-0.75em\raise-1.1ex\hbox{$\sim\;$}}}
\def\lsim{\raise0.3ex\hbox{$\;<$\kern-0.75em\raise-1.1ex\hbox{$\sim\;$}}}

\newcommand{\ba}[1]{\begin{eqnarray} \label{(#1)}}
\newcommand{\ea}{\end{eqnarray}}

\def\gsim{\raise0.3ex\hbox{$\;>$\kern-0.75em\raise-1.1ex\hbox{$\sim\;$}}}
\def\lsim{\raise0.3ex\hbox{$\;<$\kern-0.75em\raise-1.1ex\hbox{$\sim\;$}}}

\newcommand{\vtext}[1]{\begin{sideways}#1\end{sideways}}

\makeatletter
\g@addto@macro\bfseries{\boldmath}
\newcommand\Label[1]{&\refstepcounter{equation}(\mathrm{\theequation})\ltx@label{#1}&}
\makeatother

\graphicspath{{figs/}}

\allowdisplaybreaks

\title{Heavy neutral leptons and top quarks in effective field theory}

\author[a]{Rebeca Beltr\'an\,,} %\orcidlink{0000-0001-7615-8478},}
\emailAdd{rebeca.beltran@ific.uv.es}
\affiliation[a]{AHEP Group, Instituto de F\'{\i}sica Corpuscular --
  CSIC/Universitat de Val{\`e}ncia, Apartado 22085,
  E--46071 Val{\`e}ncia, Spain}

\author[b,c]{Giovanna Cottin\,,}%\orcidlink{0000-0002-5308-5808},}
\emailAdd{gfcottin@uc.cl}
\affiliation[b]{Instituto de F\'isica, Pontificia Universidad Cat\'olica de Chile, Avenida Vicu\~{n}a Mackenna 4860, Santiago, Chile}
\affiliation[c]{Millennium Institute for Subatomic Physics at the High Energy Frontier (SAPHIR), Fern\'andez Concha 700, Santiago, Chile}

\author[d]{Julian G\"unther\,,}%\orcidlink{0000-0003-3233-4753},}
\emailAdd{guenther@physik.uni-bonn.de}
\affiliation[d]{Bethe Center for Theoretical Physics \& Physikalisches Institut der 
	Universit\"at Bonn,\\ Nu{\ss}allee 12, 
	53115 Bonn, Germany}

\author[a]{Martin Hirsch\,,}%\orcidlink{0000-0001-6843-804X},}
\emailAdd{mahirsch@ific.uv.es}

\author[e]{Arsenii Titov\,,}%\orcidlink{0000-0003-1311-6072},}
\emailAdd{arsenii.titov@df.unipi.it}
\affiliation[e]{Dipartimento di Fisica ``Enrico Fermi'', 
Università di Pisa and INFN, Sezione di Pisa, \\
Largo Bruno Pontecorvo 3, I--56127 Pisa, Italy}

\author[f]{Zeren~Simon~Wang\,}%\orcidlink{0000-0002-1483-6314}}
\emailAdd{wzs@hfut.edu.cn}
\affiliation[f]{School of Physics, Hefei University of Technology, Hefei 230601, China}

\abstract{We study the phenomenology of heavy neutral leptons (HNLs) at the LHC in effective field theory, concentrating on $d=6$ operators with top quarks.
Depending on the operator choice and HNL mass, the HNLs will be produced either from proton-proton collisions in association with a single top, or via non-standard decays of top quarks.
For long-lived HNLs we estimate the sensitivity reach of different detectors to various operators with top quarks and the HNLs for the high-luminosity phase of the LHC.
For certain operators, ATLAS and some far detectors (MATHUSLA and ANUBIS) will be able to probe the associated new-physics scale as large as 12~TeV and 4.5~TeV, respectively, covering complementary HNL-mass ranges.
}

\begin{document}
\maketitle

%%%%%%%%%%%%%%%%%%%%%%%%%%%%%%%%%%%%%%%%%%%%%%%%%%%%%%%%%%%%%%%%%%%%%%
%\tableofcontents
%
%%%%%%%%%%%%%%%%%%%%%%%%%%%%%%%%%%%%%%%%%%%%%%%%%%%%%%%%%%%%%%%%%%%%%%

%\newpage

\input{subtex/01_intro.tex}

\input{subtex/02_model.tex}

\input{subtex/03_exp_simulation.tex}
\input{subtex/04_results.tex}
\input{subtex/05_conclusions.tex}

%\bigskip
\section*{Acknowledgements}

%\medskip

We thank Juan Carlos Helo for useful discussions.
This work is supported by the Spanish grants PID2023-147306NB-I00 and CEX2023-001292-S (MCIU/AEI/10.13039/501100011033), as well as CIPROM/2021/054 (Generalitat Valenciana).
R.B. acknowledges financial support from the Generalitat Valenciana (grant ACIF/2021/052). G.C.~acknowledges support from ANID FONDECYT grant No.~1250135 and ANID – Millennium Science Initiative Program ICN2019\_044.

% Bibliography
\bibliographystyle{JHEP}
\bibliography{top_hnl}

\end{document}

%% file: subtex/01_intro.tex
% !TEX root = ../top_hnl.tex
\section{Introduction}
\label{sec:intro}
%%%%%%%%%%%%%%%%%%%%%%%%%%%%%

The long-lived-particle (LLP) programs at the LHC have gained considerable momentum in the past few years~\cite{Curtin:2018mvb,Lee:2018pag,Alimena:2019zri,Feng:2022inv}.
Among the many proposals for the LLPs, one of the simplest and best motivated candidates are heavy neutral leptons (HNLs).
In minimal models, the HNLs are characterized solely by their mass and three mixing parameters (per HNL generation).
Such minimal HNLs have been searched for in many experiments.
For a recent summary of bounds, see \textit{e.g.}~Refs.~\cite{Bolton:2019pcu,Bolton:2022pyf}.
At the LHC, HNLs can be constrained particularly strongly using displaced searches, and sensitivity projections for the LHC in the high-luminosity phase can be found in Refs.~\cite{Cottin:2018kmq, Cottin:2018nms, Drewes:2019fou, Bondarenko:2019tss,Liu:2019ayx}.

New physics, however, may not be limited to a minimal HNL.
A number of UV-complete models have been discussed in the literature where the HNLs have non-minimal interactions that greatly change the expectations for HNL production (and decay) at the LHC.
Examples of such models include $Z'$ models~\cite{Deppisch:2019kvs,Chiang:2019ajm}, models with additional scalars~\cite{Deppisch:2018eth,Amrith:2018yfb}, and leptoquark models~\cite{Dorsner:2016wpm,Cottin:2021tfo}.

However, in the absence of any new resonances at the LHC, effective field theory (EFT) is probably the best tool of choice.
$N_R$-extended Standard Model effective field theory ($N_R$SMEFT) is the EFT that describes the Standard Model (SM) with additional ($n$ copies) of a light fermionic singlet~\cite{delAguila:2008ir, Aparici:2009fh,Liao:2016qyd,Bell:2005kz,Graesser:2007yj,Graesser:2007pc}.\footnote{Sometimes this is also called $\nu$SMEFT in the literature. We prefer $N_R$SMEFT to stress the (mostly) singlet character of the fermion and avoid confusions with the SM neutrinos.}
A complete classification of mass dimension $d\leq7$ operators containing $N_R$ can be found in Ref.~\cite{Liao:2016qyd}, 
and that of $d\leq9$ operators in Ref.~\cite{Li:2021tsq}, while a list of possible tree-level completions for $N_R$SMEFT operators of $d\leq7$ is given in Ref.~\cite{Beltran:2023ymm}.

A number of papers have studied the phenomenology of HNLs in the $N_R$SMEFT (or its low-energy variant $N_R$-extended low-energy EFT, $N_R$LEFT~\cite{Bischer:2019ttk,Chala:2020vqp,Li:2020lba,Li:2020wxi}) recently.
For instance, $\tau$-lepton decays to sterile neutrinos in the EFT at Belle~II were studied in Ref.~\cite{Zhou:2021ylt}.
The HNL production via $d=5$ operators in Higgs decays was considered in Refs.~\cite{Caputo:2017pit,Jones-Perez:2019plk,Barducci:2020icf,Delgado:2022fea,Duarte:2023tdw} 
and via $d=5$ and $d=6$ operators in Ref.~\cite{Butterworth:2019iff}.
Considering the HNLs produced in decays of mesons that will be copiously produced at the high-luminosity LHC, the sensitivity to $d=6$ $N_R$ operators with charged leptons was calculated in Ref.~\cite{DeVries:2020jbs} and that to $N_R$ operators with active neutrinos was obtained in Ref.~\cite{Beltran:2022ast}.
At the LHC, the HNLs in the $N_R$SMEFT can be produced also directly from parton (quark) collisions.
For this case, operators with pairs of $N_R$'s have been studied in Ref.~\cite{Cottin:2021lzz}, while for operators that produce a single $N_R$, see Refs.~\cite{Beltran:2021hpq,Mitra:2024ebr}.
We also mention Refs.~\cite{Beltran:2023nli,Fernandez-Martinez:2023phj}, which derive limits on various EFT operators with $N_R$'s from reinterpretation of previous HNL searches. 
Moreover, a set of dipole operators coupling the HNLs to the SM gauge bosons can lead to displaced decays of the HNL into a photon and a neutrino, and this scenario has been studied for several collider experiments~\cite{Barducci:2022gdv,Liu:2023nxi,Barducci:2023hzo,Barducci:2024kig,Beltran:2024twr,Barducci:2024nvd,Bertuzzo:2024eds}.
The connection between the $N_R$SMEFT and neutrinoless double beta decays with light sterile neutrinos has also been investigated in, \textit{e.g.}~Ref.~\cite{Dekens:2020ttz,DeVries:2020jbs}.
Further studies investigated the phenomenology of some of the $N_R$SMEFT operators at future lepton~\cite{Duarte:2018kiv,Zapata:2022qwo} and lepton-hadron \cite{Duarte:2014zea,Duarte:2018xst,Zapata:2023wsz} colliders (see also Refs.~\cite{Mitra:2022nri,Duarte:2025zrg}).

None of the above papers, however, considered operators with top quarks.
$N_R$'s produced from non-standard decays of the top quarks were studied in Ref.~\cite{Alcaide:2019pnf}, for the particular case of collider-stable (massless or nearly massless) $N_R$.
Since for a stable $N_R$ the final state of the top quark decay is very similar to
%standard SM top decay
that in the SM, the background is rather large and the limits that could potentially be obtained are quite weak: Ref.~\cite{Alcaide:2019pnf} estimates a limit on the scale $\Lambda \ge 0.33$ TeV for an integrated luminosity of ${\cal L}=3$~ab$^{-1}$. 
Recently, Ref.~\cite{Bahl:2023xkw} discussed production and decay rates of (long-lived) SM singlets including $N_R$, induced by four-fermion operators 
involving the top quark. However, no detector simulations were performed therein.

In this paper, we will study long-lived HNLs produced in the framework of the $N_R$SMEFT either from direct production in association with a top quark, or from top quark decays.
The HNL is assumed to be long-lived enough to either produce a displaced signal in the ATLAS or CMS detector, or even lead to signal events in one of the proposed ``far'' detectors including MATHUSLA~\cite{Chou:2016lxi} and ANUBIS~\cite{Bauer:2019vqk}.
The phenomenology depends on the operator type under consideration.
We distinguish pair and singly produced $N_R$.
For pair-$N_R$ operators, the $N_R$ can decay only via active-sterile-neutrino mixing, whereas for operators with a single $N_R$ both production and decay can be induced by the same operator.
In the latter case, it then depends strongly on the mass of the HNL, to determine whether the $N_R$ decay is dominated by the mixing or the operator.
Different from Ref.~\cite{Alcaide:2019pnf}, the displaced vertex from the $N_R$ decay can be used to reduce backgrounds.
Thus, the sensitivity to $N_R$SMEFT operator scales is greatly improved.
We estimate that with ${\cal L}=3$ ab$^{-1}$ of integrated luminosity, scales up to $\Lambda \simeq 12~(4.5)$~TeV could be probed by ATLAS (MATHUSLA and ANUBIS).%
%\textcolor{red}{{\bf More.}}
\footnote{We note that while in the present work we will focus 
on direct limits that could be set at the HL-LHC, 
it would also be interesting to study 
indirect effects of the top-$N_R$ operators and derive associated bounds, 
similarly to what has been done for the top operators in the SMEFT~\cite{Garosi:2023yxg}.}

The rest of this paper is organized as follows.
In section~\ref{sec:EFT}, we define the $d=6$ operators that we choose to study and discuss the production and decay modes of the HNL associated to these operators.
In section~\ref{sec:experiments}, we provide a brief summary of the experiments that we will consider and the detail of the numerical simulation.
In section~\ref{sec:results}, we present and discuss the sensitivity results.
We summarize our findings and conclude in section~\ref{sec:conclusions}.

%% file: subtex/02_model.tex
% !TEX root = ../top_hnl.tex
\section{Effective operators and benchmark scenarios}
\label{sec:EFT}
%%%%%%%%%%%%%%%%%%%%%%%%%%%%%
%
We assume the existence of (i) a right-handed neutrino $N_R$ with mass
$m_N$ below or around the weak scale $v = 246$~GeV, and (ii) heavy new
states at the scale $\Lambda \gg v$ that mediate interactions between
$N_R$ and the top quark $t$. $N_R$ will couple to the SM lepton
doublets $L_\ell = (\nu_{\ell L},\ell_L)^T$, $\ell=e,\mu,\tau$, and the Higgs doublet $H = (H^+,H^0)^T$ via:
\begin{equation}\label{eq:Yuk}
{\cal L}_\mathrm{Yuk} = Y^{\nu}_\ell \overline{L_\ell} \tilde{H} N_R + \text{h.c.},
%\overline{N_R}L_i\cdot H +\hskip2mm {\rm h.c.}
\end{equation}
where $\tilde{H} = \epsilon H^\ast$, 
with $\epsilon$ being the Levi-Civita symbol in two dimensions.
After electro-weak symmetry breaking, this interaction will lead
to a mixing between SM neutrinos and heavy neutral leptons,
$V_{\ell N}$. Note that, without specifying the nature of the HNL, it is
not possible to relate $V_{\ell N}$ (or, equivalently, the Yukawa
coupling) to the active neutrino masses. We will treat $V_{\ell N}$ as a
free parameter. The HNL mass could be either of Majorana type (as,
for example, in the classical type-I seesaw) or of Dirac type. There
are some subtle differences between the two cases, which affect
slightly our results. Below, for definiteness, we will show results
for a Dirac HNL. Towards the end of section \ref{sec:results}, we
will briefly comment on the changes of our results for the Majorana case.

At energies much smaller than $\Lambda$, interactions of heavy
resonances can be parameterized by higher-dimensional operators
involving $N_R$ and $t$, which are invariant under the SM gauge
symmetry; see \textit{e.g.}~Ref.~\cite{Alcaide:2019pnf}.  In this
work, we focus on the lepton- and baryon-number-conserving
four-fermion operators with these fields.  We assume one generation of
$N_R$ and, for simplicity, consider the first and third generations of
the SM quarks, and the first generation of SM leptons.  The
interactions of interest can be divided into pair-$N_R$ and
single-$N_R$ operators and are shown in
table~\ref{tab:topNops}. Note that, for simplicity, we do not use
  second generation quark indices. We comment in passing that our
  results for the production cross sections would not change
  significantly if also second generation quark indices were switched
  on.
\begin{table}[t]
\centering 
\renewcommand{\arraystretch}{1.5}
\begin{tabular}[t]{|c|l|l|}
  \hline
  & Name & Structure (+ h.c. when needed) \\ 
  \hline
  \hline
  \multirow{4}{*}{\vtext{Pair-$N_R$}} & ${\cal O}_{uN}^{13}$ &
$\boldsymbol{\left(\overline{u_R}\gamma^{\mu}t_R\right)\left(\overline{N_R}\gamma_{\mu}N_R\right)}$ \\ 
 & ${\cal O}_{uN}^{33}$ &
$\left(\overline{t_R}\gamma^{\mu}t_R\right)\left(\overline{N_R}\gamma_{\mu}N_R\right)$ \\ 
  \cline{2-3}
  & ${\cal O}_{QN}^{13}$ &
  $ \left(\overline{Q_1}\gamma^{\mu}Q_3\right)\left(\overline{N_R}\gamma_{\mu}N_R\right) = \boldsymbol{\left(\overline{u_L}\gamma^{\mu}t_L\right)\left(\overline{N_R}\gamma_{\mu}N_R\right)}+ \left(\overline{d_L}\gamma^{\mu}b_L\right)\left(\overline{N_R}\gamma_{\mu}N_R\right)$ \\ 
  & ${\cal O}_{QN}^{33}$ &
  $ \left(\overline{Q_3}\gamma^{\mu}Q_3\right)\left(\overline{N_R}\gamma_{\mu}N_R\right) = \left(\overline{t_L}\gamma^{\mu}t_L\right)\left(\overline{N_R}\gamma_{\mu}N_R\right)+ \left(\overline{b_L}\gamma^{\mu}b_L\right)\left(\overline{N_R}\gamma_{\mu}N_R\right)$ \\
  \hline 
  \hline
  \multirow{9}{*}{\vtext{Single-$N_R$}} & ${\cal O}_{duNe}^{13}$ &
$\boldsymbol{\left(\overline{d_R}\gamma^{\mu}t_R\right)\left(\overline{N_R}\gamma_{\mu}e_R\right)}$ \\ 
& ${\cal O}_{duNe}^{33}$ &
$\boldsymbol{\left(\overline{b_R}\gamma^{\mu}t_R\right)\left(\overline{N_R}\gamma_{\mu}e_R\right)}$ \\ 
  \cline{2-3}
 & ${\cal O}_{LNQd}^{31}$ &
 $\left(\overline{L}N_R\right)\epsilon\left(\overline{Q_3}^T d_R\right) 
 = \left(\overline{\nu_L}N_R\right)\left(\overline{b_L}d_R\right) 
 - \boldsymbol{\left(\overline{e_L}N_R\right)\left(\overline{t_L}d_R\right)}$ \\
 & ${\cal O}_{LNQd}^{33}$ &
 $\left(\overline{L}N_R\right)\epsilon\left(\overline{Q_3}^T b_R\right) 
 = \left(\overline{\nu_L}N_R\right)\left(\overline{b_L}b_R\right) 
 - \boldsymbol{\left(\overline{e_L}N_R\right)\left(\overline{t_L}b_R\right)}$ \\
 \cline{2-3}
 & ${\cal O}_{LdQN}^{13}$ &
 $\left(\overline{L}d_R\right)\epsilon\left(\overline{Q_3}^T N_R\right) 
 = \left(\overline{\nu_L}d_R\right)\left(\overline{b_L}N_R\right) 
 - \boldsymbol{\left(\overline{e_L}d_R\right)\left(\overline{t_L}N_R\right)}$ \\
 & ${\cal O}_{LdQN}^{33}$ &
 $\left(\overline{L}b_R\right)\epsilon\left(\overline{Q_3}^T N_R\right) 
 = \left(\overline{\nu_L}b_R\right)\left(\overline{b_L}N_R\right) 
 - \boldsymbol{\left(\overline{e_L}b_R\right)\left(\overline{t_L}N_R\right)}$ \\
 \cline{2-3}
 & ${\cal O}_{QuNL}^{13}$ &
 $\left(\overline{Q_1}t_R\right)\left(\overline{N_R}L\right) 
 = \boldsymbol{\left(\overline{u_L}t_R\right)\left(\overline{N_R}\nu_L\right) }
 + \boldsymbol{\left(\overline{d_L}t_R\right)\left(\overline{N_R}e_L\right)}$ \\   
 & ${\cal O}_{QuNL}^{31}$ &
 $\left(\overline{Q_3}u_R\right)\left(\overline{N_R}L\right) 
 = \boldsymbol{\left(\overline{t_L}u_R\right)\left(\overline{N_R}\nu_L\right) }
 + \left(\overline{b_L}u_R\right)\left(\overline{N_R}e_L\right)$ \\
   & ${\cal O}_{QuNL}^{33}$ &
 $\left(\overline{Q_3}t_R\right)\left(\overline{N_R}L\right) 
 = \left(\overline{t_L}t_R\right)\left(\overline{N_R}\nu_L\right) 
 + \boldsymbol{\left(\overline{b_L}t_R\right)\left(\overline{N_R}e_L\right)}$ \\
  \hline
\end{tabular}\\
 \caption{Lepton- and baryon-number-conserving four-fermion pair-$N_R$
   and single-$N_R$ operators with the top quark. The indices label
   the quark generations.  We focus on the first and third quark
   generations.  The terms in boldface are relevant for the HNL
   production and/or decay associated with one top quark.  }
 \label{tab:topNops}
\end{table}
%
%Here
In the table, $Q_i = (u_{iL},d_{iL})^T$ and $L=(\nu_L,e_L)^T$ denote the
SU$(2)_L$ doublets. %and $\epsilon$ is the Levi-Civita symbol in two dimensions. 
We note that the operator
\begin{align*}
 {\cal O}_{LdQN}^{i3} &= \left(\overline{L}d_{iR}\right) \epsilon \left(\overline{Q_3}^TN_R\right) 
 = \left(\overline{\nu_L}d_{iR}\right)\left(\overline{b_L}N_R\right) 
  - \left(\overline{e_L}d_{iR}\right)\left(\overline{t_L}N_R\right) \\
 &= - \frac{1}{2} {\cal O}_{LNQd}^{3i} 
 - \frac{1}{8} \left[\left(\overline{\nu_L}\sigma^{\mu\nu}N_{R}\right)\left(\overline{b_L}\sigma_{\mu\nu}d_{iR}\right) 
 - \left(\overline{e_L}\sigma^{\mu\nu}N_{R}\right)\left(\overline{t_L}\sigma_{\mu\nu}d_{iR}\right)\right],
\end{align*}
where $\sigma^{\mu\nu} = i [\gamma^\mu,\gamma^\nu]/2$, has the same
particle content as ${\cal O}_{LNQd}^{3i}$, but includes a different
Lorentz structure.

%%%%%%%%%%%%%%%%%%%%%%%%%%%%%
\subsection{HNL production}
\label{sec:HNLproduction}
%%%%%%%%%%%%%%%%%%%%%%%%%%%%%
%
\paragraph{Pair-$N_R$ operators.}

\begin{figure}[t]
\centering
\includegraphics[width=0.8\textwidth]{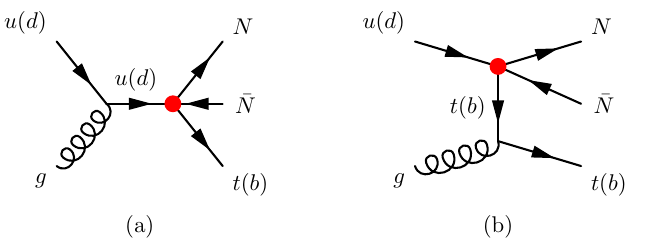}
\caption{Production of an $N\bar{N}$-pair in association with a $t$-quark 
($t$- or $b$-quark) through ${\cal O}_{uN}^{13}$ (${\cal O}_{QN}^{13}$) 
denoted by the red blob. 
(The diagrams with the particles in parentheses are present for ${\cal O}_{QN}^{13}$, but not for ${\cal O}_{uN}^{13}$.)}
\label{fig:single-top_pair-N}
\end{figure}

The pair-$N_R$ interactions ${\cal O}_{uN}^{13}$ and ${\cal O}_{QN}^{13}$
induce the flavor-violating rare top decay $t \to u N \bar{N}$.  The
corresponding decay width is given by \cite{Alcaide:2019pnf}
\footnote{In Ref.~\cite{Alcaide:2019pnf}, $m_N$ is assumed to be
negligibly small. In what follows, we keep track of it, since we are
interested in a broad range of the HNL masses.}
\begin{equation}
 \Gamma(t \rightarrow u N \bar{N}) = \frac{m_t^5 g(x)}{1536\pi^3\Lambda^4}
\left[\left(c_{uN}^{13}\right)^2 + \left(c_{QN}^{13}\right)^2\right],
\end{equation}
where
\begin{align}
 g(x) &= \left(1-14x-2x^2-12x^3\right) \sqrt{1-4x} \nonumber \\
 &\phantom{{}={}}-12x^2\left(1-x^2\right) \left[\ln\frac{1-\sqrt{1-4x}}{1+\sqrt{1-4x}} 
 -\ln\frac{1+\sqrt{1-4x}-2x}{2x}\right],
\end{align}
and $x = m_N^2/m_t^2$.
At the same time, these operators mediate the direct production of 
an $N\bar{N}$-pair in association with a top quark 
(as well as with a bottom quark for ${\cal O}_{QN}^{13}$)
in $pp$ collisions at the LHC, 
through the diagrams shown in figure~\ref{fig:single-top_pair-N}.

At the LHC, top quarks are dominantly produced in pair through the strong interaction. 
The inclusive cross-section of $t\bar{t}$ production at the LHC 
with $\sqrt{s}=13$~TeV is 
$\sigma_{t\bar{t}} = 830 \pm 38$~pb~\cite{ATLAS:2020aln}. 
The single-top production through the weak interaction 
has a smaller cross-section: $\sigma_{tq+\bar{t}q} = 221 \pm 13$~pb at $\sqrt{s}=13$~TeV~\cite{ATLAS:2024ojr}. 
In what follows, when considering the SM production of the top quark,
we will focus on the top quark pair production. 

\begin{figure}[t]
\centering
\includegraphics[width=0.48\textwidth]{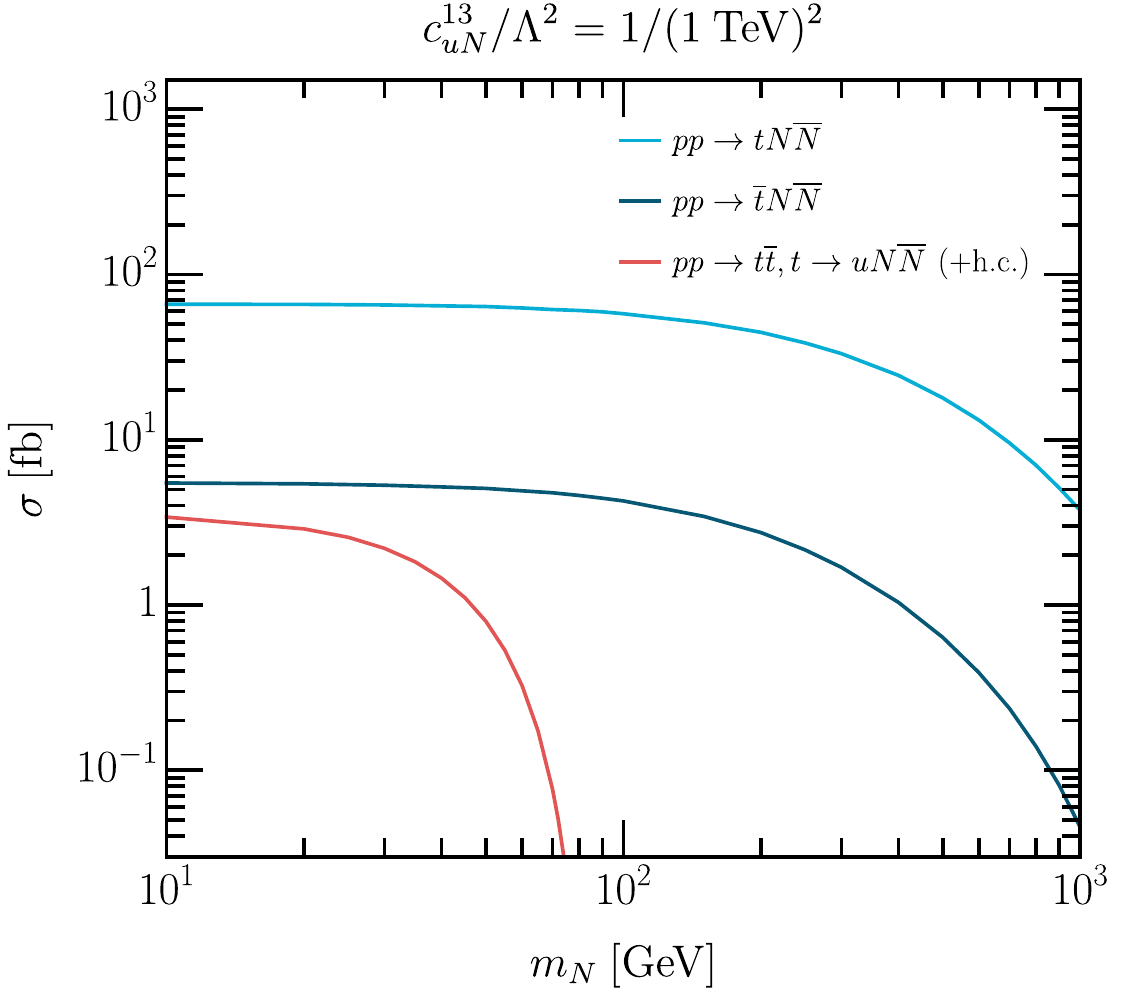}
\includegraphics[width=0.48\textwidth]{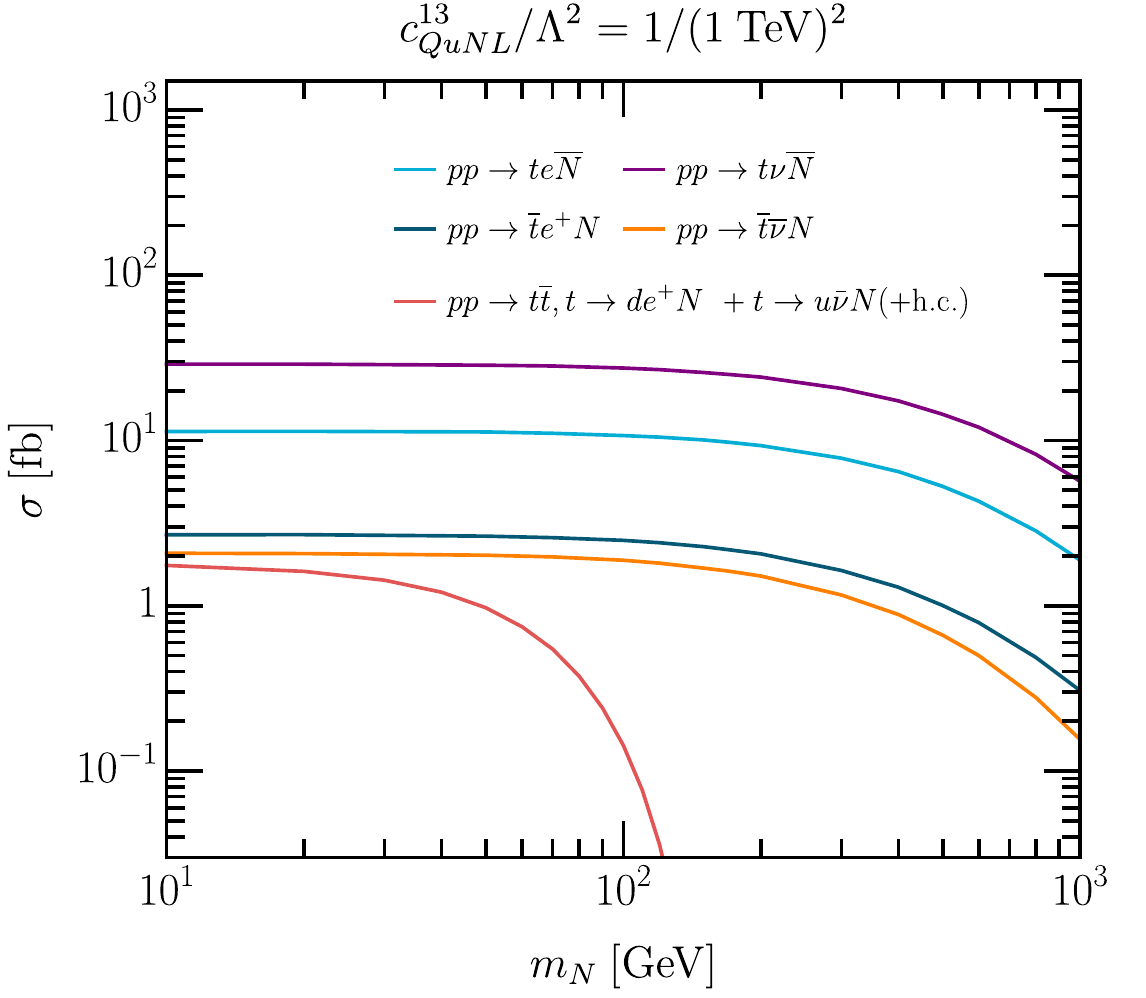}

\includegraphics[width=0.48\textwidth]{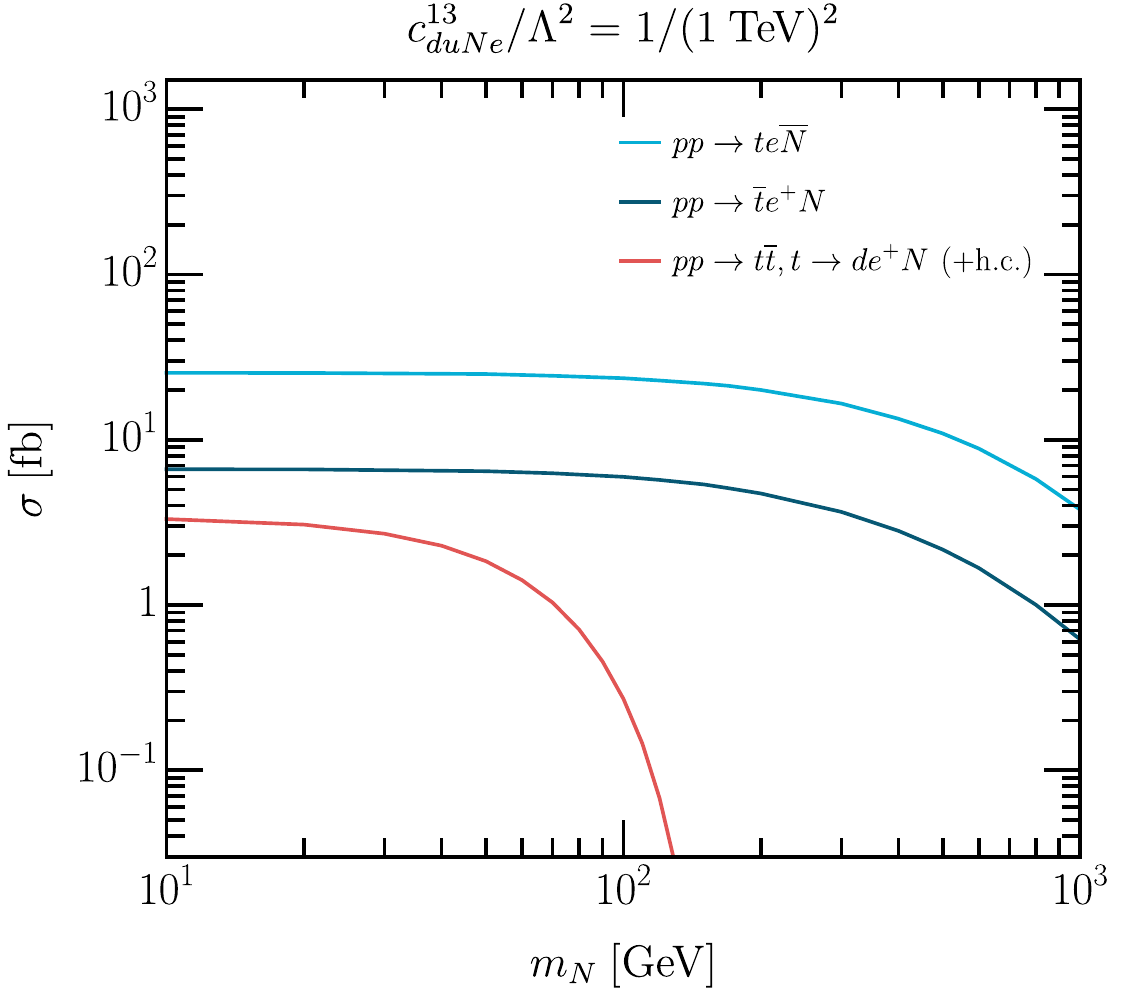}
\includegraphics[width=0.48\textwidth]{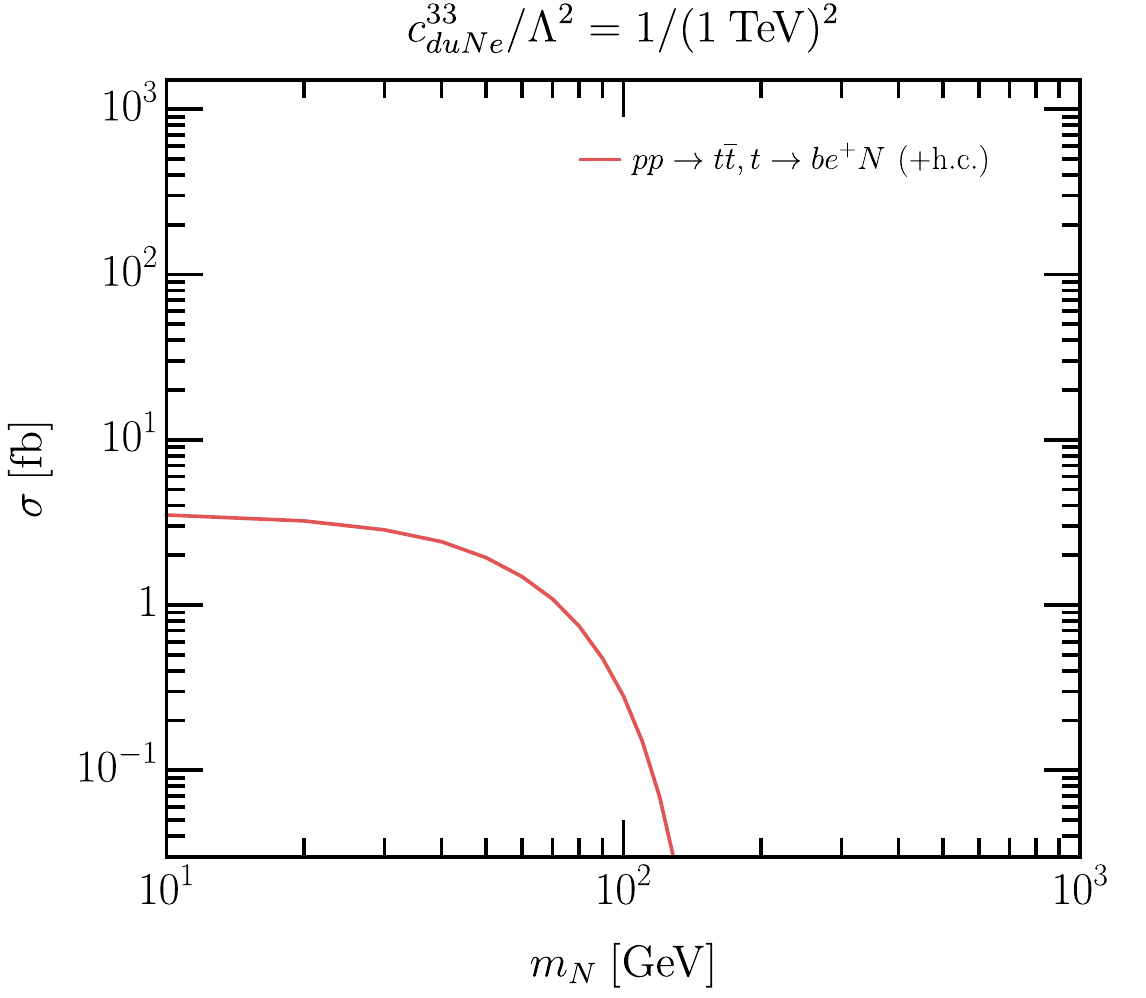}

\caption{HNL production cross-sections induced by the operators 
${\cal O}_{uN}^{13}$ (top left), ${\cal O}_{QuNL}^{13}$ (top right), ${\cal O}_{duNe}^{13}$ (bottom left), ${\cal O}_{duNe}^{33}$ (bottom right), for $c_\mathcal{O}^{ij} = 1$ and $\Lambda = 1$~TeV at the LHC with $\sqrt{s}=14$~TeV.
}
\label{fig:Xsecs}
\end{figure}

To give an example of the relative importance of the two HNL production mechanisms introduced above, we simulate with \texttt{MadGraph5}~\cite{Alwall:2011uj,Alwall:2014hca} (i)~$pp \to t\bar{t}$ with a subsequent rare top decay $t \to u N \bar{N}$, (ii)~$pp \to t N \bar{N}$, and (iii)~$pp \to \bar{t} N \bar{N}$ induced by the operator $\mathcal{O}_{uN}^{13}$.
The resulting cross-sections obtained 
for $\Lambda = 1$~TeV and $\sqrt{s} = 14$~TeV 
are shown in the upper-left panel of figure~\ref{fig:Xsecs}.
The individual production modes are identified by labels in the plots.
We observe that $\sigma(pp \to tN\bar{N})$ and $\sigma(pp \to \bar{t}N\bar{N})$ 
dominate over $\sigma(pp \to t\bar{t}) [\mathcal{B}(t \to u N \bar{N}) + \mathcal{B}(\bar{t} \to \bar{u} N \bar{N})]$ 
%\AT{is the sum what you plot?}
in the whole range of HNL masses of interest, where $\mathcal{B}$ denotes decay branching ratio.\footnote{We generate events at leading order and multiply only the top-antitop quark production cross-section by a flat factor of $k \sim 1.7$ corresponding to the cross-section determined at NNLO+NNLL by the \texttt{Top++2.0} program~\cite{Czakon:2011xx}. We do not introduce correction factors for HNL production rate in direct $pp$ collisions.}
The main difference between the shape of the cross-sections for these two production modes is the behavior at larger $m_N$ values. The contribution from top decays is highly suppressed when $m_N$ approaches $m_t/2$, whereas in direct $pp$ collisions the cross-section remains relatively flat up to a few hundreds of GeV.
The difference in the cross-sections for $pp \to tN\bar{N}$ and $pp \to \bar{t}N\bar{N}$ is due to the larger parton distribution function (PDF) 
of the $u$-quark compared to that of $\bar{u}$ inside the proton.
We note that the production cross-sections for the states $tN{\bar N}$/${\bar t}N{\bar N}$ from operator ${\cal O}_{QN}^{13}$ are identical to the ones shown for ${\cal O}_{uN}^{13}$.
We therefore do not show plots for this case.

The operators ${\cal O}_{uN}^{33}$ and ${\cal O}_{QN}^{33}$ neither trigger a top decay, nor do they contribute to the direct HNL production in $pp$ collisions.
Therefore, we will not consider them in what follows.

\paragraph{Single-$N_R$ operators.}
The operators with off-diagonal quark-flavor indices, \textit{viz.}~13 and 31, 
trigger the flavor-violating decays $t \to d e^+ N$ 
and $t \to u \bar{\nu} N/ u \nu \bar{N}$
with the following widths:~\cite{Alcaide:2019pnf}
\begin{align}
 \Gamma(t \to d e^+ N) &= \frac{m_t^5 f(x)}{6144 \pi^3 \Lambda^4} 
 %\left( 1 - 8x + 8x^3 -x^4 -12x^2 \ln{x} \right) \nonumber \\
 %&\times
 \bigg[4\left(c_{duNe}^{13}\right)^2 
 + \left(c_{QuNL}^{13}\right)^2 \nonumber \\
 &\hspace{2.5cm}+ \left(c_{LNQd}^{31}\right)^2 
 + \left(c_{LdQN}^{13}\right)^2 - c_{LNQd}^{31}c_{LdQN}^{13}\bigg]\,,  \label{eq:tTOdeN} \\
 \Gamma(t \to u \bar{\nu} N) &= \frac{m_t^5 f(x)}{6144 \pi^3 \Lambda^4} \left(c_{QuNL}^{13}\right)^2
 \quad \text{and} \quad 
 \Gamma(t \to u \nu \bar{N}) = \frac{m_t^5 f(x)}{6144 \pi^3 \Lambda^4} \left(c_{QuNL}^{31}\right)^2\,,
 \label{eq:tTOuvN}
\end{align} 
where $f(x) = 1 - 8x + 8x^3 -x^4 -12x^2 \ln{x}$, and $x = m_N^2/m_t^2$.
The same operators also contribute to the single-top production in association with either a charged lepton and an HNL, or an SM neutrino and an HNL, directly in $pp$ collisions.

The operators with $Q_3=(t_L,b_L)^T$, \textit{i.e.}~${\cal O}_{LNQd}^{31}$, ${\cal O}_{LdQN}^{13}$, and ${\cal O}_{QuNL}^{31}$, in addition give a contribution to the single-bottom production in association with either $\nu\bar{N}$ or $e^+ N$. 

\begin{figure}[t]
\centering
\includegraphics[width=0.8\textwidth]{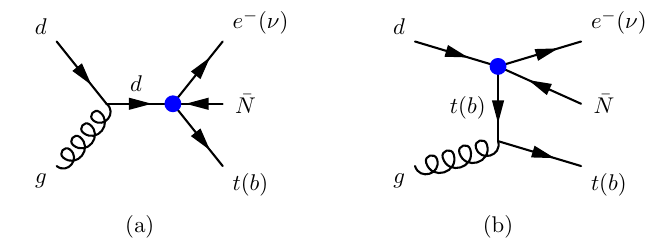}
\caption{HNL production in association with $e^-$ and $t$ 
($e^-$ and $t$ or $\nu$ and $b$) through ${\cal O}_{duNe}^{13}$ (${\cal O}_{LNQd}^{31}$) 
denoted by the blue blob. 
The diagrams with the particles in parentheses are present for ${\cal O}_{LNQd}^{31}$, but not for ${\cal O}_{duNe}^{13}$.}
\label{fig:single-top_single-N}
\end{figure}

As an example, we show in figure~\ref{fig:single-top_single-N} 
the diagrams for the processes triggered by ${\cal O}_{duNe}^{13}$ 
and ${\cal O}_{LNQd}^{31}$.

In the lower-left panel of figure~\ref{fig:Xsecs}, we display the cross-sections for (i)~$pp \to t\bar{t}$ followed by a subsequent decay $t \to de^+ N$, (ii)~$pp \to t e^- \bar{N}$, and (iii)~$pp \to \bar{t} e^+ N$, triggered by ${\cal O}_{duNe}^{13}$, setting $\Lambda = 1$~TeV. 
As in the case of the pair-$N_R$ operator ${\cal O}_{uN}^{13}$, 
the direct HNL production in $pp$ collisions dominates 
over the production through the new top quark decay. 
Compared to ${\cal O}_{uN}^{13}$, the cross-section 
for $pp \to t e^- \bar{N}$ is smaller than that for 
$pp \to t N \bar{N}$, which is due to a smaller PDF of the $d$-quark with respect to the one of the $u$-quark.

In the upper-right panel of figure~\ref{fig:Xsecs}, we show the cross-sections for the processes induced by $\mathcal{O}_{QuNL}^{13}$. 
Since this operator involves SU(2)$_L$ doublets $Q_1$ and $L$, we have both $te\bar{N}$ and $t\nu\bar{N}$ final states (as well as their charge conjugates).
As in the previous cases, the direct HNL production in $pp$ collisions dominates over the production through new top quark decays.

The operators with the 33 quark-flavor indices lead to the flavor-conserving decay $t \to b e^+ N$ for which the width can be computed by eq.~\eqref{eq:tTOdeN} with all quark-flavor indices set to 33.%
\footnote{ Note that constraints on these operators  derived from the measured top quark width are very weak, namely, $\Lambda \gsim 0.14$ TeV~\cite{Biswas:2024gtr},  which is actually below the scale of EFT validity. Thus, all our study points fulfil this limit automatically.}
The contribution of these flavor-diagonal operators to the single-top and single-bottom production (in association with an electron 
or neutrino and an HNL) is negligible because of the smallness of the $b$-quark PDF.
As an example, we show in the lower-right panel of figure~\ref{fig:Xsecs} the only HNL production process that is induced by $\mathcal{O}_{duNe}^{33}$. 

In table~\ref{tab:Nproddec}, we summarize the HNL production modes and decay channels for each effective operator of interest.
We discuss the latter in the next subsection.
\begin{table}[t]
\centering
\renewcommand*{\arraystretch}{1.5}
\begin{tabular}{|c|c|c|c|c|c|}
\hline 
\multicolumn{2}{|c|}{\textbf{Operator}} & \multicolumn{2}{c|}{\textbf{HNL production modes}} & \multicolumn{2}{c|}{\textbf{HNL decay channels}} \\ \hline
Name & Flavor & Top decay & $pp$ collision & 5-body & 3-body \\ 
\hline
\hline
$\mathcal{O}_{uN}$ & 13 & 
$t \to u N \bar{N}$ & $pp \to t N \bar{N} $ & 
$\times$ & $\times$ \\ 
\hline
$\mathcal{O}_{QN}$ & 13 & 
$t \to u N \bar{N}$ & $pp \to t N \bar{N} / b N \bar{N} $ & 
$\times$ & $\times$ \\ 
\hline
\hline
\multirow{3}{*}{$\mathcal{O}_{QuNL}$} & 13 & 
$t \to u \bar{\nu} N / d e^+ N$ & $ p p  \to t \nu \bar{N} / t e^- \bar{N}$ &
$N \to \nu\, t^\ast \bar{u} / e^- t^\ast \bar{d} $ & $\times$ \\
& 31 & 
$ t \to  u \nu \bar{N}$ & $ p p \to t \bar{\nu} N / b e^+ N $ & 
$N \to \nu\, u\, \bar{t}^\ast $ & $N \to e^- u \bar{b}$ \\
& 33 & 
$t \to b e^+ N$ & $\times$ & 
$N \to  e^- t^\ast \bar{b} / \nu t^\ast \bar{t}^\ast $ & $\times$ \\
\hline
\multirow{2}{*}{$\mathcal{O}_{duNe}$} & 13 & 
$t \to d e^+ N$ & $ p p \to t e^- \bar{N}$ & 
$N \to e^- t^\ast \bar{d}$ & $\times$ \\ 
& 33 & 
$t \to  b e^+ N$ & $\times$ & 
$N \to e^- t^\ast \bar{b}$ & $\times$ \\
\hline
\multirow{2}{*}{$\mathcal{O}_{LNQd}$} & 31 & 
$t \to d e^+ N$ & $ p p \to t e^- \bar{N} / b \nu \bar{N}$ & 
$N \to e^- t^\ast \bar{d} $ & $N \to \nu b \bar{d}$ \\ 
& 33 & 
$t \to b e^+ N$ & $\times$ & 
$N \to e^- t^\ast \bar{b} $ & $N \to \nu b \bar{b}$ \\
\hline
\end{tabular}
\caption{HNL production models and tree-level decay channels induced by the four-fermion operators with $N_R$ and third-generation quarks. 
Pair-$N_R$ operators contribute only to HNL production, 
while single-$N_R$ operators also induce 
5-body and in some cases 3-body HNL decays. 
The virtual top quark $t^\ast$ decays as $t^\ast \to b W^{\ast} \to b (q' \bar{q}~\text{or}~\ell^+\nu)$. 
The operators ${\cal O}_{LdQN}^{13}$ and ${\cal O}_{LdQN}^{33}$ lead to the same processes as ${\cal O}_{LNQd}^{31}$ and ${\cal O}_{LNQd}^{33}$, respectively.
}
\label{tab:Nproddec}
\end{table}
%

%%%%%%%%%%%%%%%%%%%%%%%%%%%%%
\subsection{HNL decays}
\label{sec:HNLdecay}
%%%%%%%%%%%%%%%%%%%%%%%%%%%%%
%
\paragraph{Pair-$N_R$ operators.} 
The pair-$N_R$ interactions cannot make the HNL decay. 
Hence, in these scenarios, the $N$ decay can proceed only via active-sterile-neutrino mixing. 
In this case, we use the formulae for the HNL decay width provided in Ref.~\cite{Bondarenko:2018ptm}.

\begin{figure}[t]
 \centering
 \includegraphics[width=0.8\textwidth]{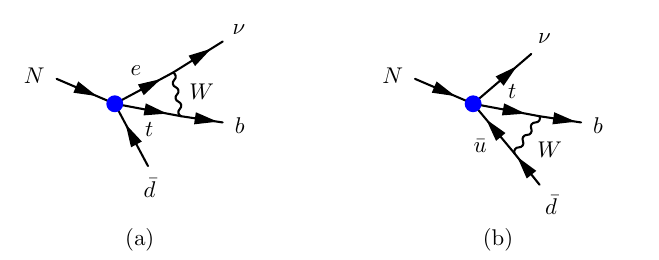}
 \caption{One-loop diagrams (in the unitary gauge) 
 for the 3-body decay $N\to\nu b \bar{d}$ induced by the operator
 $\mathcal{O}_{QuNL}^{13}$.}
 \label{fig:1loopQuNL13}
\end{figure}

\paragraph{Single-$N_R$ operators.}
These operators lead to HNL decays even in the absence of active-sterile-neutrino mixing.
We discuss the corresponding decay modes below.
\begin{itemize}
  \item All single-$N_R$ operators induce 5-body decays of $N$ through an off-shell top quark, $t^\ast$, and an off-shell $W$-boson, $W^\ast$.
  For example, the decay chain triggered by 
  $\mathcal{O}_{duNe}^{33}$ is 
  $N \to e^- t^* \overline{b} \to e^- b W^{*} \overline{b}  \to e^- b\overline{b}\, (q'\overline{q}~\text{or}~\ell^+ \nu_\ell)$.% 
  \footnote{If $m_N$ is large enough, this decay becomes either $4$- or $3$-body, with respectively an on-shell $W$ or an on-shell $t$ in the final state.} 
  Analogous decay chains for the other single-$N_R$ operators are given in table~\ref{tab:Nproddec}.
 \item Operators containing two terms, 
 one of which does not involve the top quark, 
trigger 3-body decays.% 
\footnote{We assume that $m_N$ is sufficiently large 
for $b$-quark(s) to go on shell.}
There are five flavor structures in this category: 
$\mathcal{O}_{LNQd}^{31}$, $\mathcal{O}_{LNQd}^{33}$,
$\mathcal{O}_{LdQN}^{13}$, $\mathcal{O}_{LdQN}^{33}$,
 and $\mathcal{O}_{QuNL}^{31}$, \textit{cf}.~table~\ref{tab:topNops}. 
  For example, in the case of $\mathcal{O}_{LNQd}^{33}$, 
  we have $N \to \nu b \overline{b}$. 
  The corresponding 3-body decays for the other operator structures are shown in table~\ref{tab:Nproddec}.
  \item At the one-loop level, the operators leading to 5-body tree-level decays
  will also trigger 3-body decays. The corresponding decay rates 
  can be sizable for the operators with a particular chiral structure. 
  For example, the operators $\mathcal{O}_{QuNL}^{13}$ and $\mathcal{O}_{QuNL}^{33}$ contain light left-handed fields together with a right-handed top quark $t_R$, 
  which can be easily turned to a left-handed top quark~$t_L$ by the large top Yukawa coupling
  and coupled to the $W$-boson~\cite{Bahl:2023xkw}.
  The corresponding one-loop diagrams are shown in figure~\ref{fig:1loopQuNL13} 
  for the case of $\mathcal{O}_{QuNL}^{13}$.
\end{itemize}
The relative contributions of these decay modes to the total HNL decay width will depend on the model parameters, \textit{i.e.}~the HNL mass, $m_N$, and the interaction couplings, $c_{\mathcal{O}}/\Lambda^2$ and $V_{e N}$. 
To give an example, in what follows we consider three operator structures, $\mathcal{O}_{duNe}^{33}$, $\mathcal{O}_{LNQd}^{33}$, and $\mathcal{O}_{QuNL}^{13}$, and compare their contributions to the $N$ decay width with that in the minimal scenario, characterized by active-sterile-neutrino mixing only.

\begin{figure}[t]
\centering
\includegraphics[width=0.49\textwidth]{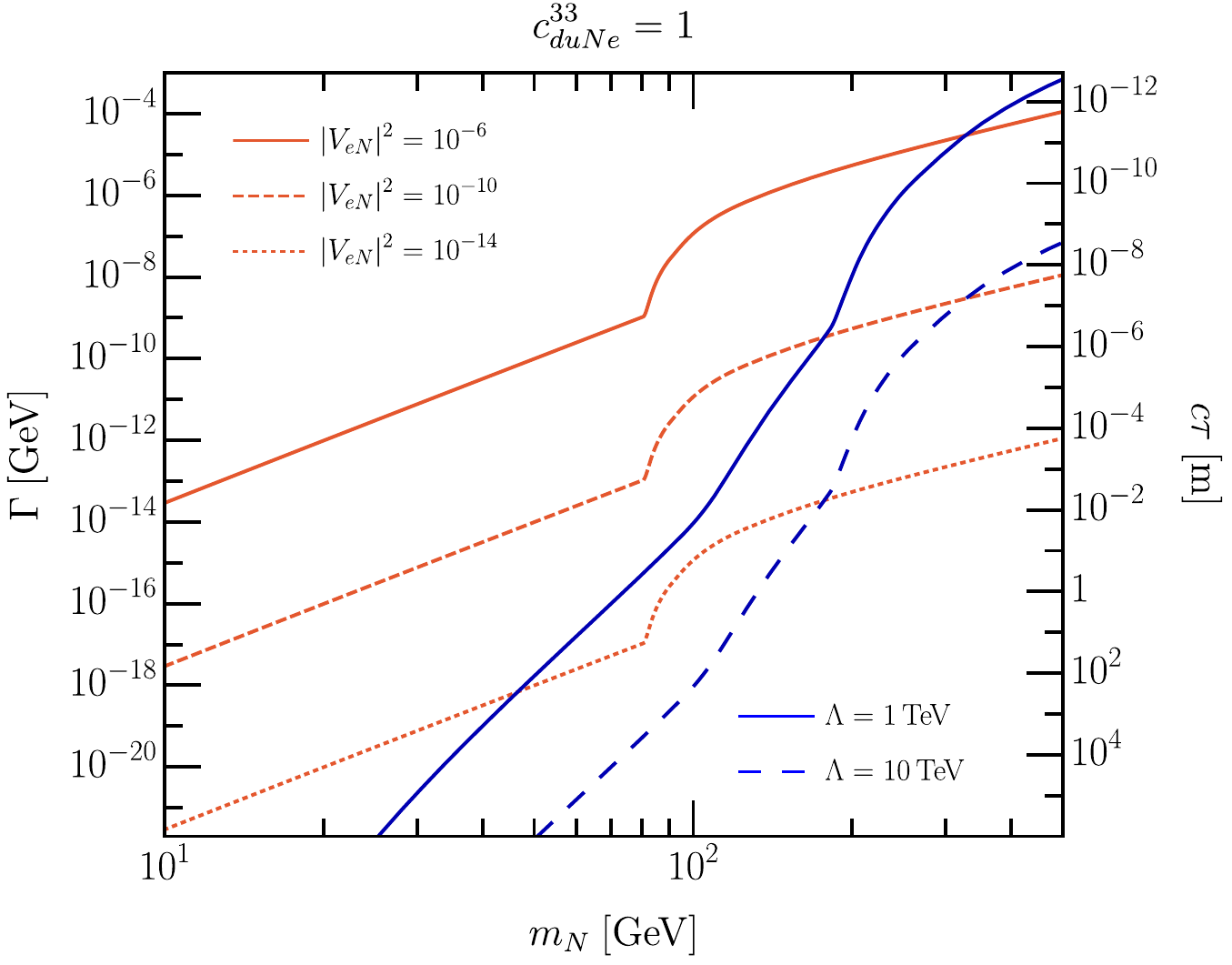}
\hfill
\includegraphics[width=0.49\textwidth]{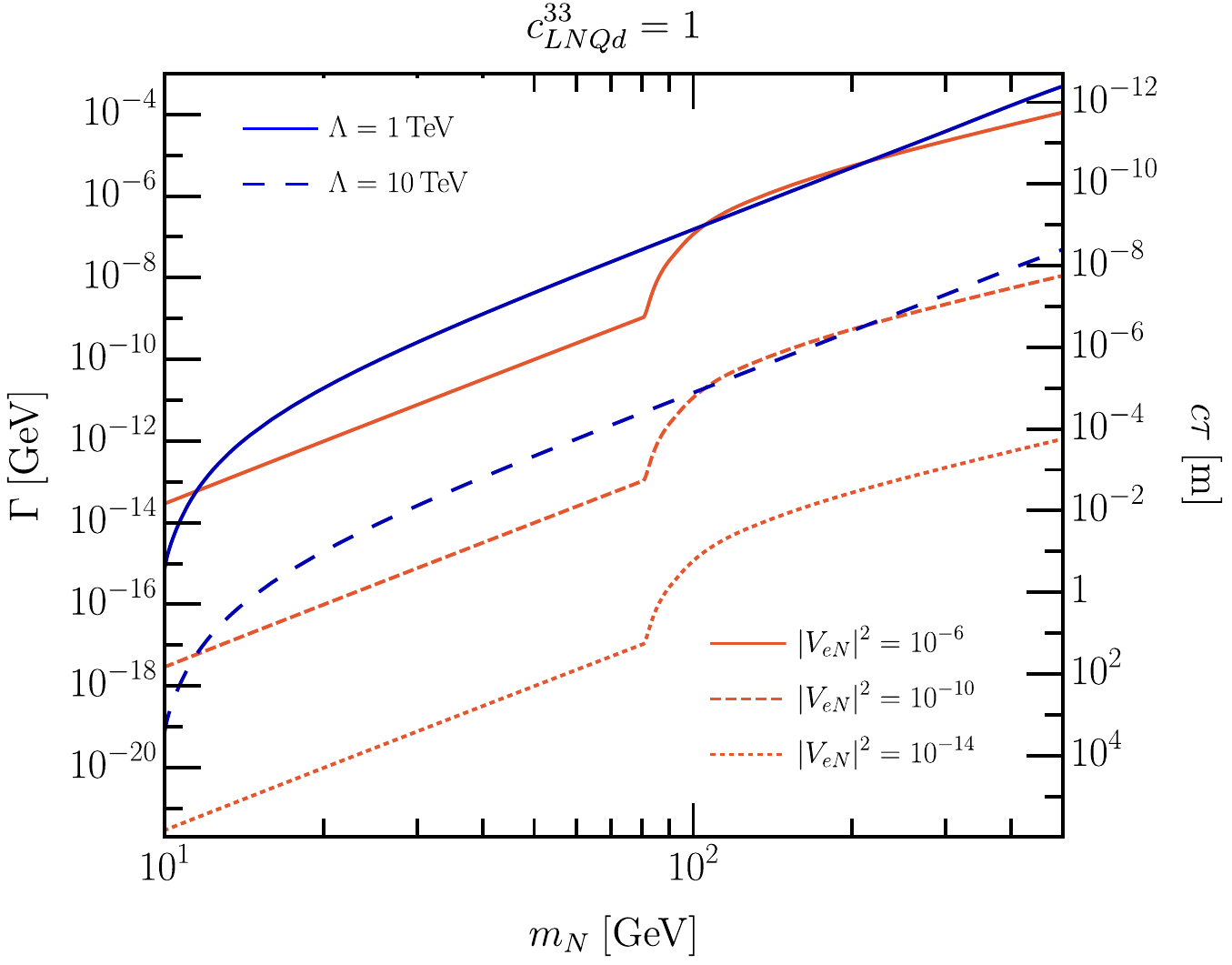}\\[0.1cm]
\includegraphics[width=0.49\textwidth]{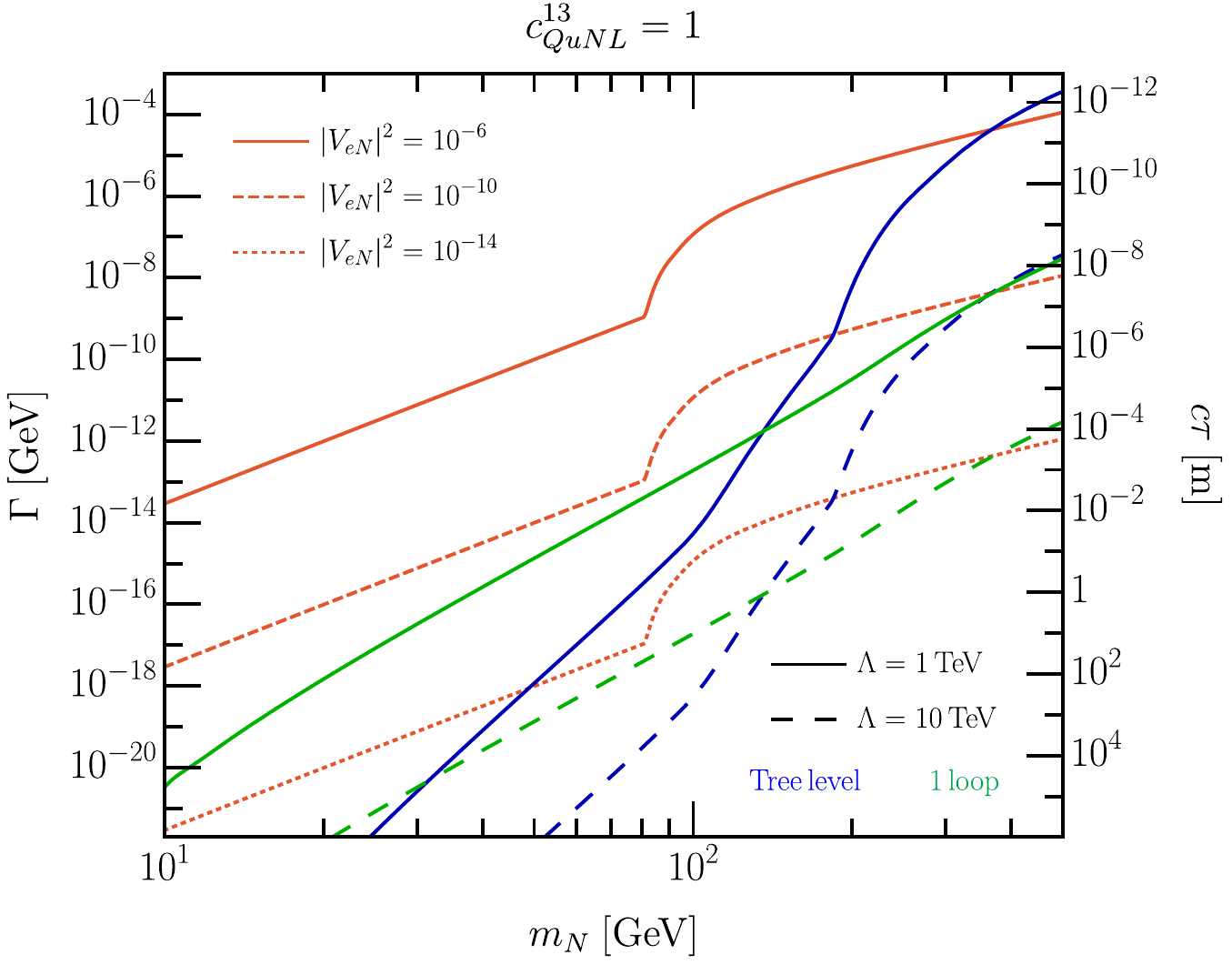}
\caption{Mixing and operator contributions to the HNL decay width as a function of its mass.
	The operator contributions (blue and green lines) come from $\mathcal{O}_{duNe}^{33}$ (top left), $\mathcal{O}_{LNQd}^{33}$ (top right), and $\mathcal{O}_{QuNL}^{13}$ (bottom), 
and we have set two values of $\Lambda$. 
Contributions from three values of $|V_{e N}|^2$ are shown for comparison in orange.
}
\label{fig:widths}
\end{figure}
In the upper-left panel of figure~\ref{fig:widths}, 
we display the HNL partial decay widths induced by $\mathcal{O}_{duNe}^{33}$, namely, 
$\Gamma (N \to e^- b\bar{b} q'\bar{q}) + \Gamma (N \to e^- b\bar{b} \ell^+\nu_\ell) $, where $q'\bar{q} = \{ u\bar{d}, u\bar{s}, c\bar{s}, c\bar{d}\}$ 
and $\ell^+ \nu_\ell = \{e^+ \nu_e, \mu^+ \nu_\mu, \tau^+ \nu_\tau\}$.%
\footnote{The decays to $q'\bar{q}= \{u\bar{b}, c\bar{b}\}$ are strongly suppressed by the corresponding small CKM elements. 
By summing over all possible final states we are taking into account the contributions from off-shell and on-shell top quark and $W$-boson.}  We have set $c_{duNe}^{33}/\Lambda^2 = 1 / (1 \text{ TeV})^2$. For comparison, we show the mixing contribution to the decay width for three different values of the squared mixing parameter $|V_{eN}|^2$. 
Notice the $y$-axis spans around 20 orders of magnitude along the mass range under consideration. There are two sudden increases in the operator contribution, which occur at $m_N \simeq m_W + 2 m_b$ and $m_N \simeq m_t + m_b$, as the decays to on-shell $W$-bosons and top quarks, respectively, become possible. 
A similar behavior can be observed in the mixing curves at $m_N \simeq m_W$, when the decay channel to on-shell $W$ opens.

In the upper-right panel of figure~\ref{fig:widths}, we show the HNL partial decay width of the 3-body decay $N \to \nu b \bar{b}$ triggered by the operator $\mathcal{O}_{LNQd}^{33}$ (it dominates over the 5-body decays, which we do not show). We have set $c_{LNQd}^{33} /\Lambda^2 = 1/(1\text{ TeV})^2$.
The operator contribution wins over the mixing one if $\Lambda$ is not too large, 
as $\Gamma_{\cal O} \propto \Lambda^{-4}$ 
and $\Gamma_\text{mix} \propto |V_{e N}|^2$. 

In the bottom panel of figure~\ref{fig:widths}, by the green lines 
we show the decay width 
of the 3-body decay $N \to \nu b \bar{d}$ induced by $\mathcal{O}_{QuNL}^{13}$ through the one-loop diagrams depicted in figure~\ref{fig:1loopQuNL13}. 
We have computed this decay width using the pipeline \texttt{FeynArts}~\cite{Hahn:2000kx} + \texttt{FormCalc}~\cite{Hahn:1998yk} + \texttt{LoopTools}~\cite{Hahn:1998yk}. 
We find perfect agreement with the corresponding result shown in figure~6 of Ref.~\cite{Bahl:2023xkw}. 
The blue lines correspond to the sum of the ``5-body'' tree-level decay widths for $N \to e^- t^* \bar{d} \to e^- W^* b \bar{d} \to e^{-} b \bar{d} (q'\bar{q}~\text{or}~\ell^+\nu_\ell)$ 
and 
$N \to \nu_e t^* \bar{u} \to \nu_e W^* b \bar{u} \to \nu_e b \bar{u} (q'\bar{q}~\text{or}~\ell^+\nu_\ell)$, induced by $\mathcal{O}_{QuNL}^{13}$ at tree-level 
and computed with \texttt{MadGraph5}. 
As can be seen, for $c_{QuNL}^{13}/\Lambda^2 = 1/(1~\text{TeV})^2$, 
the loop-induced 3-body decay dominates over the tree-level 5-body decays 
for HNL masses up to approximately $m_N\approx130$~GeV.

In the case of $\mathcal{O}_{QuNL}^{33}$, we have a similar 3-body decay $N \to \nu b \bar{b}$.
However, now, in addition to the diagrams with the $W$-boson in the loop, we also have diagrams with exchange of the $Z$-boson and the Higgs boson. 
We have checked that for $\Lambda = 1$~TeV and $m_N \gtrsim 15$~GeV, 
this operator leads to prompt decays with $c\tau \lesssim 0.01$~mm. 
Thus, we will not consider it in what follows among our 
benchmark scenarios featuring a long-lived HNL.

%%%%%%%%%%%%%%%%%%%%%%%%%%%%%
\subsection{Benchmark scenarios}
\label{sec:benchmarks}
%%%%%%%%%%%%%%%%%%%%%%%%%%%%%
%
In section~\ref{sec:results}, we will study numerically a few
representative scenarios, each characterized by an effective operator
with given quark-flavor indices.  We will switch on one operator
structure at a time and investigate its phenomenological implications
at the LHC main detector ATLAS, as well as at the current
and future ``far detector'' facilities, including MoEDAL-MAPP2,
MATHUSLA, ANUBIS and CODEX-b.

We describe the characteristic features of each selected scenario below, 
and we refer the reader to table~\ref{tab:Nproddec} for a summary of the HNL production and decay modes.
\begin{enumerate}
 \item ${\cal O}_{uN}^{13}$: Being a pair-$N_R$ operator, it only contributes to pair production of HNLs. The dominant channel is $pp \to t N \bar{N}$, \textit{cf.}~the upper-left panel of figure~\ref{fig:Xsecs}. HNL decay, on the other hand, is completely controlled by the mixing parameter. We note in passing that a scenario with ${\cal O}_{QN}^{13}$ would be qualitatively very similar.
 \item  ${\cal O}_{QuNL}^{13}$: This operator contributes to both HNL production and decay. As discussed above and shown in the bottom panel of figure~\ref{fig:widths}, the 3-body decay induced at one loop dominates over 5-body decays induced at tree-level.
In our numerical simulation, we will take into account both the 3-body decay via the operator and tree-level decays via mixing.

 \item ${\cal O}_{duNe}^{13}$: This operator structure also contributes to both production and decay. HNLs are dominantly produced directly in $pp$ collisions, as can be inferred from the upper-right panel of figure~\ref{fig:Xsecs}. 
 The operator contribution to the decay is naturally suppressed (compared to mixing) for $m_N< m_W$, since the only allowed channel is a 5-body decay.\footnote{This is true only at tree-level. One can consider 3-body decays via $W$-loop. 

 Their contribution to the total decay width %of this channels
 is negligible though for this particular operator structure.} 
 This is no longer true for $m_N > m_W (m_t)$, when 4-body (3-body) decays are possible.
 Therefore, the operator contribution should be taken into account along with the mixing contribution.
 \item ${\cal O}_{duNe}^{33}$: The operator contributes to HNL production and decay. However, production now occurs only through top quark decays, leading to smaller cross-sections than in the previous scenarios;  see figure~\ref{fig:Xsecs}. As for the decay, analogously to scenarios 2 and 3, the operator contribution competes with the mixing one and has to be taken into account, \textit{cf}.~the upper-left panel of figure~\ref{fig:widths} and the related discussion. 

\end{enumerate}
%

%% file: subtex/03_exp_simulation.tex
% !TEX root = ../top_hnl.tex
\section{Experiments and simulation}
\label{sec:experiments}
%%%%%%%%%%%%%%%%%%%%%%%%%%%%%

In this work, we consider not only the ATLAS experiment but also a series of proposed far detectors dedicated to LLP searches at the LHC.\footnote{We do not simulate explicitly for the CMS experiment, for which we expect similar results.}
Here, we introduce these experiments and search strategies, and outline the simulation procedure employed for determining the sensitivity reach of these experiments to long-lived HNLs coupled to top quarks.

\subsection{ATLAS}

For the ATLAS experiment, we focus on a signature characterized by a DV arising from the HNL decay, accompanied by jets originating from either top quark decays or the HNL decay itself.
The latest DV search with jets at ATLAS~\cite{ATLAS:2023oti} is not optimal for our signal, mainly because of its too strong thresholds on the transverse momenta of jets and multi-jet cuts.
Therefore, we propose a new search strategy with optimal jet cuts and the optional requirement of a $b$-jet.
Our strategy is inspired by the recast~\cite{Cheung:2024qve} for the ATLAS ``DV+jets'' search~\cite{ATLAS:2023oti} and by a past 8-TeV search for DVs at ATLAS~\cite{Aad:2015rba}.

We implement the operators of interest in a UFO file with \texttt{Feynrules}~\cite{Christensen:2008py,Alloul:2013bka}.
With this model file, we then simulate HNL production in $pp$ collisions at $\sqrt{s}=14$~TeV using \texttt{MadGraph5}~\cite{Alwall:2011uj,Alwall:2014hca} with the NNPDF3.1 PDF set~\cite{NNPDF:2017mvq}.
For each benchmark scenario, we focus on the dominant production channel of the HNL: either direct $pp$ collision in association with a top or bottom quark, or top quark decay if the former is absent (see table~\ref{tab:Nproddec}).
We generate $100$ thousand events at multiple parameter points in a grid covering the plane $|V_{eN}|^2$ vs.~$m_N$, while keeping $c_{\mathcal{O}}^{ij}/\Lambda^2$ constant for each benchmark.
HNL decays are then handled in \texttt{MadSpin}~\cite{Artoisenet:2012st}, where signal decays to final states with jets are enforced for numerical stability. In particular, we simulate the decay into $ejj$ and $\nu jj$ via mixing, and for benchmarks 2 -- 4, we additionally simulate the HNL decays induced by the single-$N_R$ operators, which all contain at least two jets.

The generated events are processed in \texttt{Pythia8}~\cite{Sjostrand:2014zea} for showering, hadronization, and event selections. We implement a toy-detector module in \texttt{Pythia8} for reconstructing truth-level jets, following the definition from the HEPData accompaying note from ref.~\cite{ATLAS:2023oti}. Truth jets are reconstructed using \texttt{FastJet}~\cite{Cacciari:2011ma}, with the anti-$k_t$ algorithm and $R = 0.4$, excluding neutrinos and muons.
This jet definition includes particles from the HNL decays. The detector response for measurement of jet $p_{T}$ is also modeled taking into account detector acceptance, resolution, and
smearing on their transverse momenta, following the same criteria and threshold as in ref.~\cite{Allanach:2016pam}. 

Event selection proceeds in two stages: (1) event-level selections; and (2) DV-level selections.

\textbf{Event-level selections:} we impose preliminary selections based on the number of jets in the event meeting specific $p_T$-criteria --- events must have 4, 5, or 6 jets with $p_T> 90,65$, or $55$~GeV, respectively, following Ref.~\cite{Aad:2015rba}.
This choice of multi-jet cuts enhances the sensitivity to our signal, as opposed to considering the jet selections imposed in the recast done in Ref.~\cite{Cheung:2024qve}.
The search~\cite{ATLAS:2023oti} that was recast in that paper was intended for directly pair-produced LLPs (decaying to several jets).
For our benchmarks, sensitivity is largely lost at the event-level when using these 13-TeV-search jet cuts~\cite{ATLAS:2023oti}, and therefore we propose to retain the jet-$p_{T}$ thresholds as low as the ATLAS 8-TeV search~\cite{Aad:2015rba}.\footnote{This proposal can be further justified by the potential and prospective leverage that some specialized data taking techniques (\textit{i.e.}~such as data parking~\cite{CMS:2024zhe}) can have on beyond the Standard Model physics on data collected with low triggering thresholds. Such techniques were used by LHC experiments in searches for HNLs; see \textit{e.g.}~Ref.~\cite{CMS:2024ita}.}

\begin{figure}[t]
	\centering
	\includegraphics[width=0.6\textwidth]{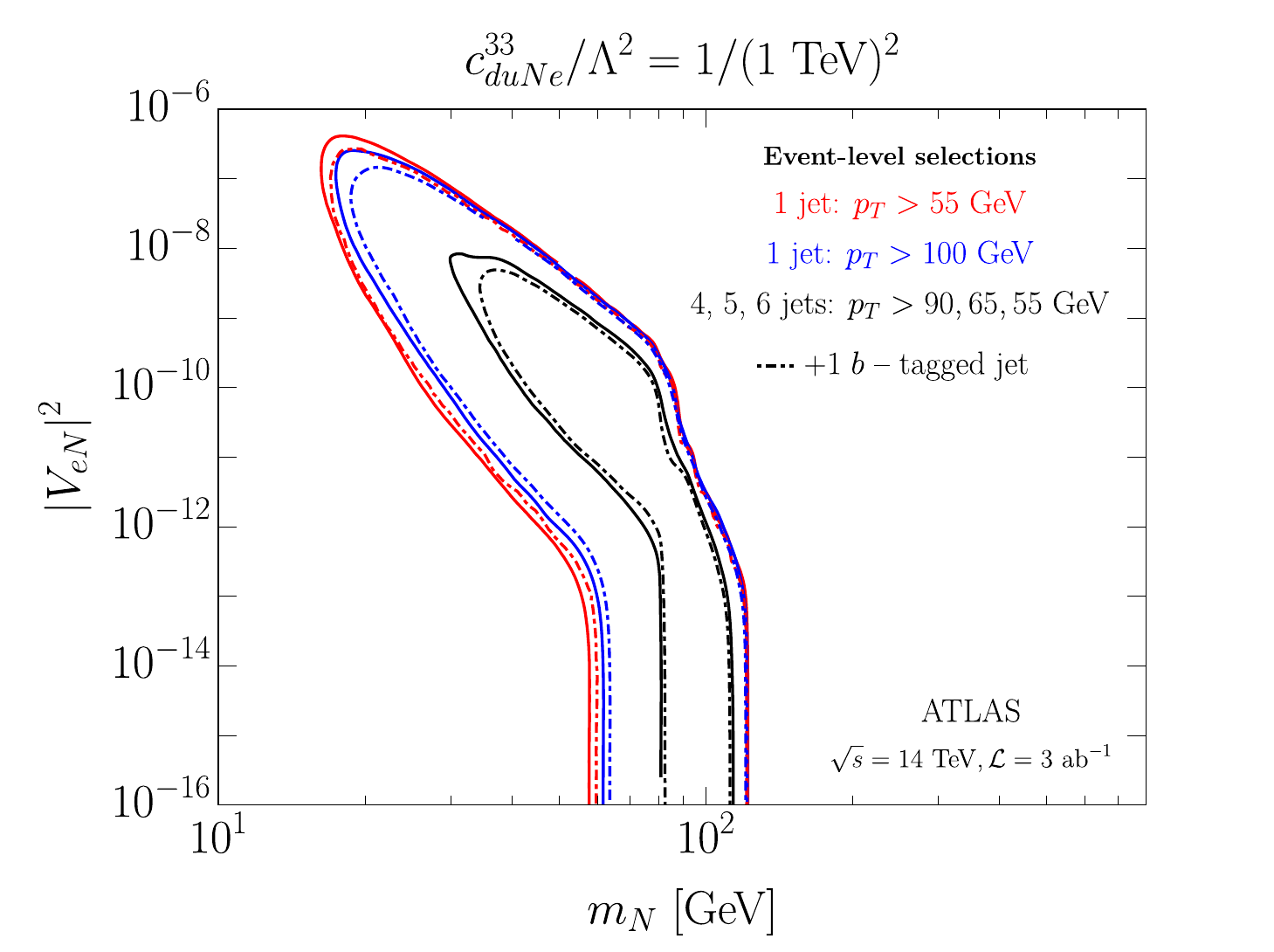}
	\caption{ATLAS sensitivity for one example operator, $\mathcal{O}_{duNe}^{33}$, in the plane $|V_{eN}|^2$ vs.~$m_N$ for the proposed complete search strategy with different jet-selection cuts at the event-level (see text for more details).
    }
	\label{fig:atlas_dune33}
\end{figure}

In figure~\ref{fig:atlas_dune33}, we show the impact on the sensitivity limits to one of our benchmarks when requiring different jet-$p_T$ thresholds.
When aiming at discovery prospects, one may further consider identifying $b$-quark jets associated to our event signatures (see table~\ref{tab:Nproddec}).
We treat a jet as a $b$-jet based on the flavor of the truth quark that initiated the jet.
From all truth jets in the event with $p_{T}>20$ GeV and $|\eta|<2.5$, we tag a jet as a $b$-jet if a Monte-Carlo truth $b$-quark with $p_{T}>5$ GeV is found within a cone of size $\Delta R=0.3$ around the jet direction.
We additionally force a flat $b$-jet identification efficiency of 77\%~\cite{ATLAS:2011qia}.

\textbf{DV-level selections:} we implement DV reconstruction according to the recast~\cite{Cheung:2024qve} of the ATLAS ``DV+jets'' search~\cite{ATLAS:2023oti}, where at least one vertex in the event must satisfy the following conditions:
\begin{enumerate}
\item The transverse distance from the vertex to the IP must be within 4~mm $< R_{xy} < 300$~mm and the longitudinal position of the vertex must satisfy $|z|<300$ mm.
\item At least one track should have an absolute transverse impact parameter $|d_0|>2$~mm.
\item The DV must contain at least 5 tracks all satisfying the two requirements below:
\begin{enumerate}
	\item They should have a boosted transverse decay length larger than 520 mm.
	\item Their $p_T$ and charge $q$ should satisfy $p_T/|q| > 1$ GeV.
\end{enumerate}
\item The invariant mass of the DV, reconstructed from the tracks passing the above requirements for which the masses are all assumed to be that of a charged pion, should be larger than 10 GeV.
\end{enumerate}

The ATLAS collaboration~\cite{ATLAS:2023oti} provided parameterized efficiencies at both event level and vertex level, to account for further, more intricate event selections that are difficult to simulate, for recasting purpose.
The event-level (vertex-level) efficiencies\footnote{The ATLAS search~\cite{ATLAS:2023oti} employs two signal regions, the strong one and the electroweak one. The two SRs share the same vertex-level efficiencies.} are functions of the transverse position of the DV and the sum of the truth-jet $p_T$ (the transverse position of the DV, the number of charged particles associated to the truth decay vertex, as well as the vertex invariant mass).
We apply the parameterized efficiencies at vertex-level only, in our proposed search.
Thus, the events that pass the jet-selection criteria and contain at least one reconstructed vertex meeting the above requirements contribute to the final signal-efficiency cutflow.

The expected number of signal events at ATLAS is calculated with
\begin{align}
N_{S}^{\text{\tiny ATLAS}} & =   \sigma \cdot \mathcal{L} \cdot \Big\{ \mathcal{B}_\text{mix}  \cdot \varepsilon^{\text{mix}} +  \mathcal{B}_\text{op} \cdot \varepsilon^{\text{op}} \Big\} \,, \\
\mathcal{B}_\text{mix}  & = \mathcal{B}(N \to ejj) +  \mathcal{B}(N \to \nu jj), \nonumber \\
 \mathcal{B}_\text{op} & = \mathcal{B}(N \to \text{5-body tree}) + \mathcal{B}(N \to \text{3-body loop})\,, \nonumber
\end{align}
where $\sigma$ is the production cross-section of the HNL $N$, $\mathcal{L}=3$ ab$^{-1}$ is the integrated luminosity, and $\mathcal{B}_{\text{mix/op}}$ represents the sum of the branching fractions of the HNL decay modes of interest induced by the mixing or the operator. We remark that the latter term is not present in benchmark 1.
Here, ``$j$" denotes a jet including the up, down, charm, strange, or bottom quark.
The final cutflow efficiencies, $\varepsilon^{\text{mix/op}}$, are the ones obtained from implementing both event- and vertex-level selections in each case (including the acceptances and parameterized efficiencies) to the final states induced via mixing/operator in our custom code implemented within \texttt{Pythia8}.

We note that the vertex-level selection criteria are designed to suppress all backgrounds,\footnote{The ATLAS DV+jets search reported in Ref.~\cite{ATLAS:2023oti} applies stricter cuts on jet-$p_T$ thresholds. However, in that search, background suppression is driven primarily by DV requirements rather than by requirements on the jets in the event, justifying our assumption that the DV selections alone are sufficient to eliminate the background events.} allowing us to derive the 95\% C.L.~exclusion limits by requiring at least 3 signal events.
We additionally provide exclusion limits for 10 and 30 signal events for certain scenarios, which would correspond to the exclusion limits at the same level if approximately 25 and 225 background events were present, respectively.

\subsection{Far detectors}
Several proposals of a far detector displaced relative to different interaction points (IPs) at the LHC have been brought up in the last decade, such as FASER(2)~\cite{Feng:2017uoz,FASER:2018eoc}, CODEX-b~\cite{Gligorov:2017nwh,Aielli:2019ivi}, MATHUSLA~\cite{Curtin:2018mvb,MATHUSLA:2020uve,Chou:2016lxi}, ANUBIS~\cite{Bauer:2019vqk}, FACET~\cite{Cerci:2021nlb}, and MoEDAL-MAPP1(2)~\cite{Pinfold:2019nqj,Pinfold:2019zwp}.
They are supposed to be detector systems with tracking capabilities enabling reconstruction of displaced vertices consisting of charged particles, to be placed about 5--500 meters from various IPs.
The LLPs, once produced \textit{e.g.}~at an IP, can travel towards a far detector and decay inside.
The macroscopic distances between these experiments and the IPs allow for implementation of efficient veto and shielding of potential background events, leading to expected vanishing background level at these experiments in general, even with an integrated luminosity as large as 3 ab$^{-1}$.
We note that FASER has been approved and its operation has been launched.

These experiments can be classified into two categories according to their relative direction with respect to their corresponding IPs, namely, transverse detectors (MATHUSLA, ANUBIS, MoEDAL-MAPP1(2), and CODEX-b) and forward detectors (FASER(2) and FACET).
In the present work, the long-lived HNLs are assumed to stem from either top quark decays or direct production in $pp$ collisions.
The HNLs thus produced tend to travel in the transverse direction.
Therefore, we will not discuss further the forward experiments which we have numerically checked and verified to have no sensitivities.
Among the transverse detectors, MATHUSLA and ANUBIS would be constructed in the vicinity of the CMS and ATLAS IPs, respectively, both with a projected integrated luminosity of 3 ab$^{-1}$ in the HL-LHC era.
The MATHUSLA experiment would be a box-shaped detector to be instrumented on the ground surface, with a 68 m (60 m) distance in the horizontal (vertical) direction from the IP.
It would have a designed geometrical size of 100 m $\times$ 100 m $\times$ 25 m.\footnote{Recently a smaller version has been suggested for cost-related reasons~\cite{mathusla_new_design}. However, since it is still in the developing stage, we stick to the latest design~\cite{MATHUSLA:2020uve}.}
ANUBIS has been proposed to be installed inside a service shaft above the ATLAS IP, with about 5 m horizontal distance in the beam direction from the IP.
It has cylindrical shape with a diameter of 18 m and a height of 54 m.\footnote{In Ref.~\cite{ANUBIS_talk_slides} a new geometrical design of ANUBIS has been proposed, where the apparatus is supposed to be instrumented on the ATLAS cavern ceiling or at the bottom of a shaft. As the design has not been finalized, we choose to focus on the original proposal.}
MoEDAL-MAPP1(2) and CODEX-b are both proposed far detectors  related to the LHCb IP8.
CODEX-b would be cubic with dimensions 10 m $\times$ 10 m $\times$ 10 m, placed 25 m from the IP.
The pseudorapidity $\eta$ and azimuthal-angle $\phi$ coverages are respectively $\eta\in [0.2,0.6]$ and $\frac{\delta\phi}{2\pi}\sim \frac{0.4}{2\pi}$.
It is projected to receive data of 300 fb$^{-1}$ at the end of the HL-LHC phase.
MoEDAL-MAPP1 and MAPP2 are proposed as trapezoidal-shaped detectors to be implemented in the UGCI gallery.
The MAPP1 detector would have a volume of 130 m$^3$ and the MAPP2 is an extended version of MAPP1, with a volume of 430 m$^3$.
MAPP1 has a polar angle of 5$^\circ$ and a distance of 55~m, relative to the IP, and MAPP2 would take up the space of the whole gallery.
For these two experiments, the expected integrated luminosities are 30 fb$^{-1}$ and 300 fb$^{-1}$, as they should be operated during the LHC Run3 and the HL-LHC periods, respectively.

For computing the sensitivity of the far detectors, we follow a somewhat different procedure than that used for ATLAS, as we do not simulate the HNL decay in this case.
We start by generating 100 thousand events for various HNL masses using \texttt{MadGraph5} and pass the resulting Les Houches Event Files (LHEF)~\cite{Alwall:2006yp} directly to \verb|Pythia8|.
We then calculate the probability for the HNL to decay within each far detector, and derive the average decay probability over all the simulated HNLs,
\begin{equation}
\langle P\left[ N \text{ decay in f.v.} \right] \rangle = \frac{1}{k} \sum^{k}_{i = 1} P \left[ N^i \text{ decay in f.v.}\right],
\end{equation}
where $k=10^5$ or $2\times 10^5$, depending on whether in each signal event one or two HNLs are produced, is the total number of the simulated HNLs for each HNL mass, ``f.v.'' stands for fiducial volume, and the probability in each case is determined by the HNL’s boosted decay length and momentum angle besides the geometry and position of the far detectors.
The expected number of signal events at the far detectors is then given by:
\begin{equation}
N_{S}^{\text{\tiny FD}} = \sigma \cdot \mathcal{L}  \cdot \langle P\left[ N \text{ decay in f.v.} \right] \rangle  \cdot \mathcal{B} (N \to \text{vis.}) \cdot \varepsilon \,,
\end{equation}
where $\sigma$ and $\mathcal{L}=3$ ab$^{-1}$ are respectively the production cross-section and the integrated luminosity of the HL-LHC era.
$\mathcal{B} (N \to \text{vis.})$ denotes the decay branching ratio of the HNL into visible final states (that can be induced either via mixing or the operator), for which we have only excluded the fully invisible tri-neutrino final states.
We assume a detection efficiency of $\varepsilon=100\%$ for simplicity.

As vanishing background is expected at the far detectors, we derive the 95\% C.L. exclusion limits by requiring 3 signal events.
However, MATHUSLA and ANUBIS might change the detector geometries (and thus the total acceptance) from the original designs, and ANUBIS with its relatively close distance from the IP might suffer from a non-zero level of background events.
Therefore, we will also provide sensitivity limits for larger numbers of signal events using the original designs, to facilitate deriving sensitivities corresponding to other values of the acceptances or background levels.

%% file: subtex/04_results.tex
% !TEX root = ../top_hnl.tex
\section{Numerical results}
\label{sec:results}
%%%%%%%%%%%%%%%%%%%%%%%%%%%%%

In this section, we present the numerical results obtained using the
simulation procedure outlined above. We analyze the sensitivities of
ATLAS and the far detectors, focusing on the four benchmark scenarios
defined in section~\ref{sec:benchmarks}, characterized by different
effective operators containing a top quark and at least one HNL. In
each case, a single effective operator structure is switched on,
together with the standard active-sterile-neutrino mixing. The mixing
parameter $|V_{\ell N}|^2$, the HNL mass $m_N$, and the operator
coefficients $c_{\mathcal{O}}^{ij}/\Lambda^2$ are treated as
independent parameters. For simplicity, we consider a single
kinematically relevant HNL that mixes exclusively with electron
neutrinos, \textit{i.e.} $V_{\ell N} = V_{eN}$.\footnote{We note that, based on a study for first generation quark operators
\cite{Beltran:2021hpq}, we expect that results for muons will be very
similar.}

We note that for the plots of the results shown in this section, we do
not assume that the bottom quarks in the event have been tagged.
Thus, our regions are, strictly speaking, not discovery regions but
exclusion prospects.  We choose to do so, to obtain a fair comparison
of the far detectors and ATLAS, because only ATLAS can identify the
top quark(s) in the event.  However, as shown in
figure~\ref{fig:atlas_dune33}, discovery regions (including
$b$-tagging) are similar to, albeit slightly smaller than, exclusion
prospects in the ATLAS simulation.

The benchmark scenarios are divided into two categories: pair-$N_R$
operators (scenario~1) and single-$N_R$ operators
(scenarios~2--4). For pair-$N_R$ operators, HNL production is
dominated by the operator, while the HNL decays occur solely via
mixing, resulting in decoupled production and decay processes. In this
case, the mixing parameter $V_{eN}$ drives leptonic and semileptonic
HNL decays through off-shell or on-shell $W$- and $Z$-bosons,
depending on the HNL mass. Conversely, for single-$N_R$ operators, the
operator contributes to both production and decay, with the latter
potentially dominating over the mixing in some parts of the parameter
space. In either case, the final states containing charged particles
are detectable by the far detectors, while the final states with jets
contribute to the searched signal in ATLAS.

\begin{figure}[t]
    \centering
        \includegraphics[width=0.49\linewidth]{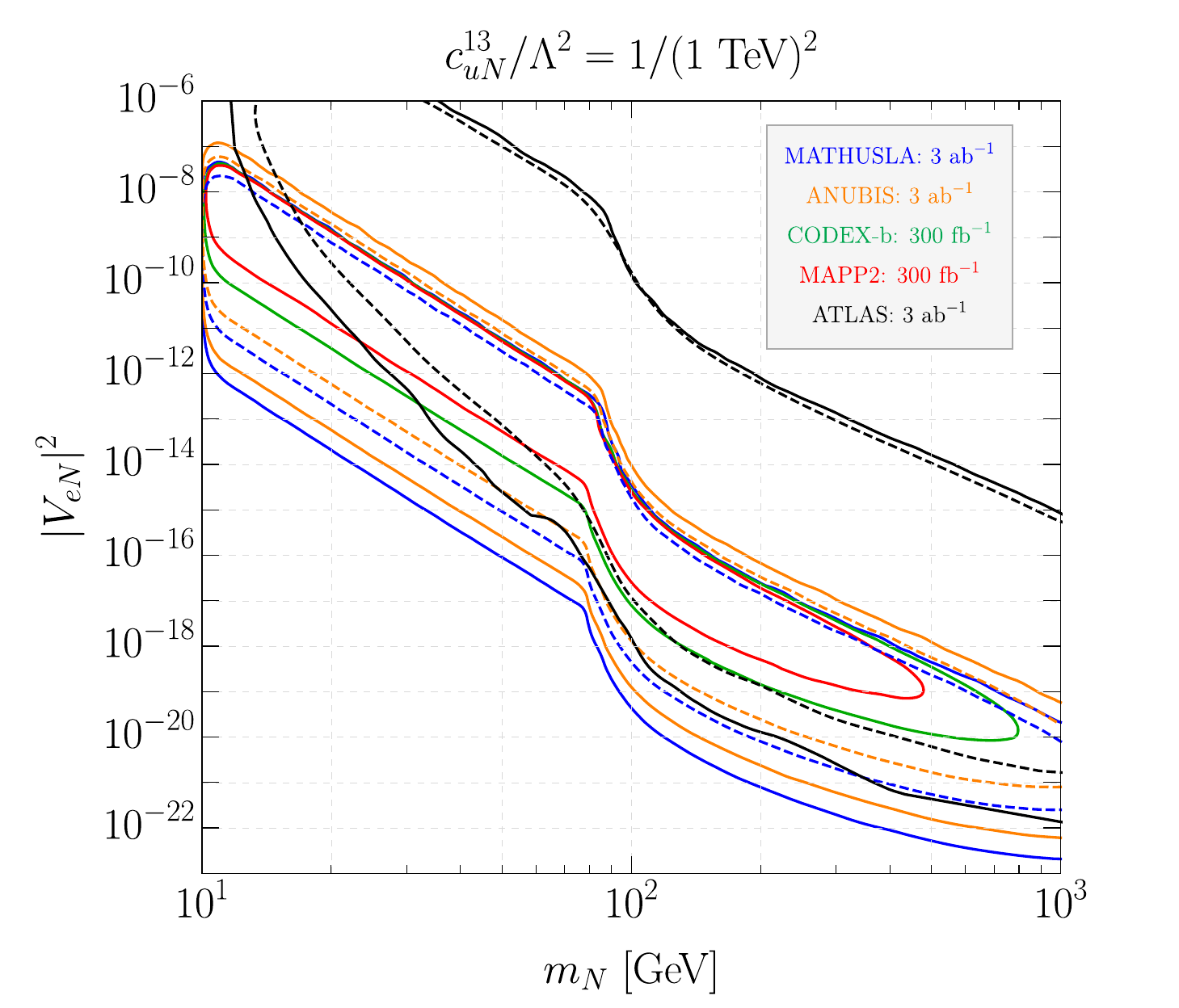} 
       \includegraphics[width=0.49\linewidth]{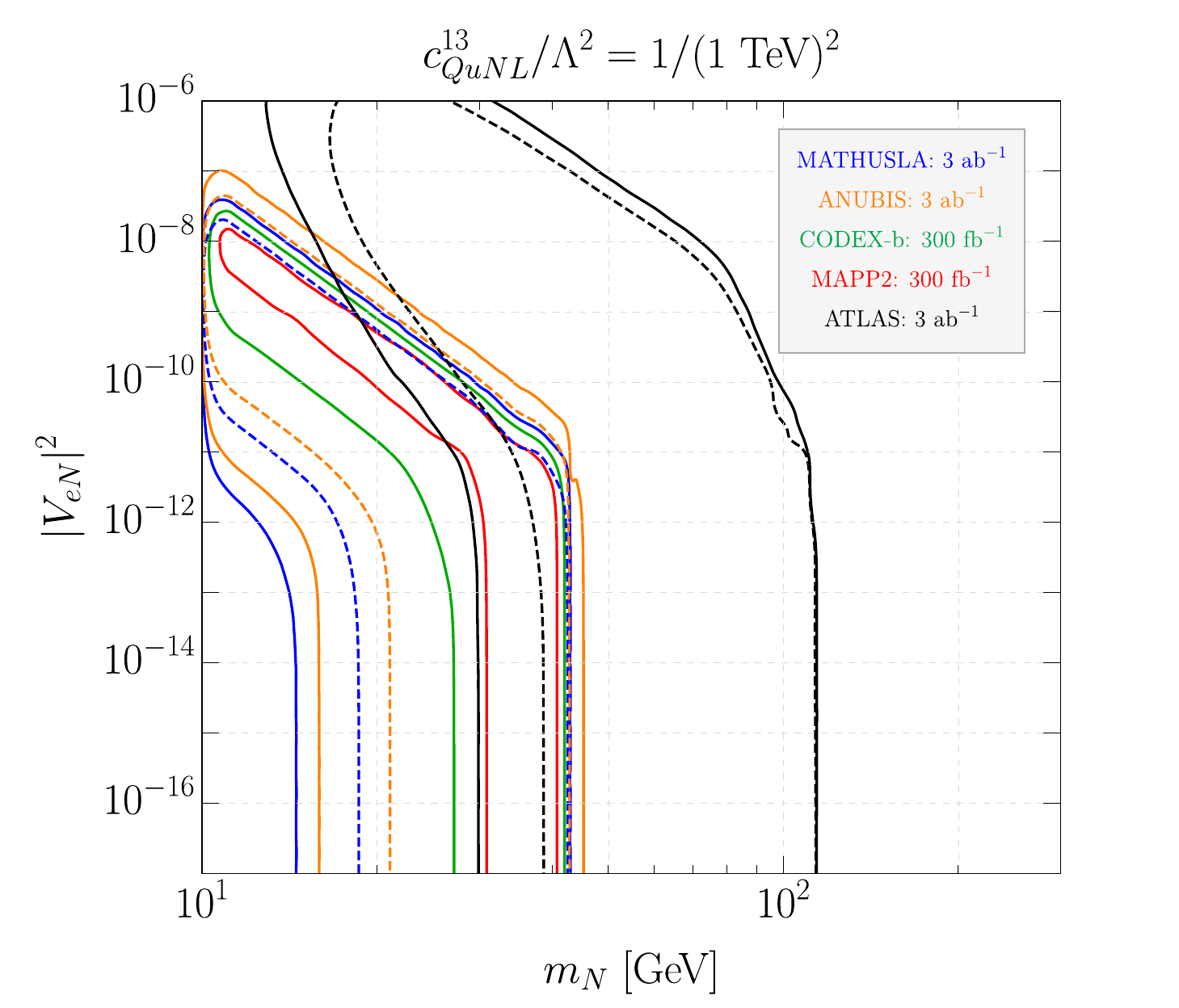} 
     \vspace{0.1cm}
         
        \includegraphics[width=0.49\linewidth]{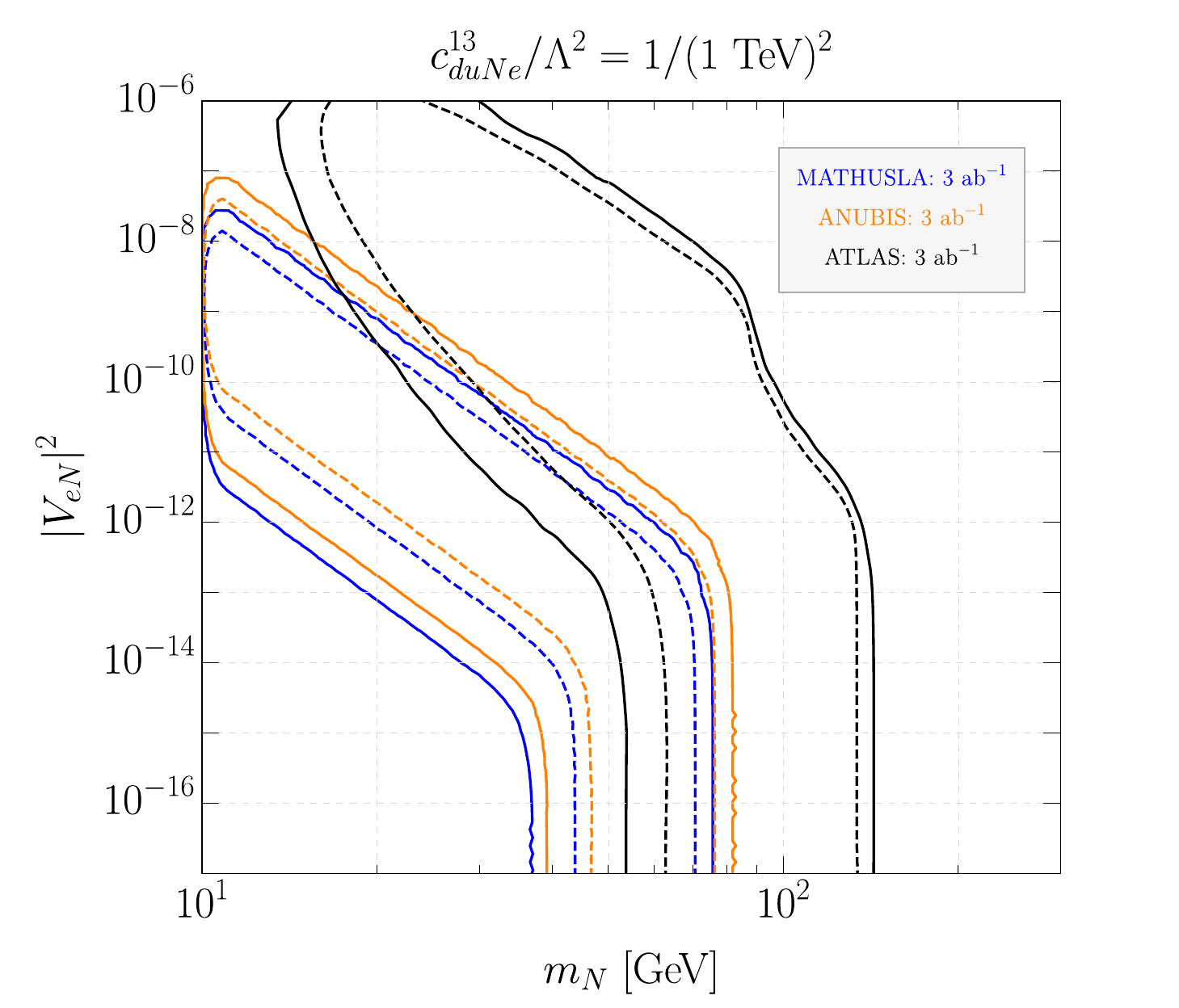} 
    \hfill
        \includegraphics[width=0.49\linewidth]{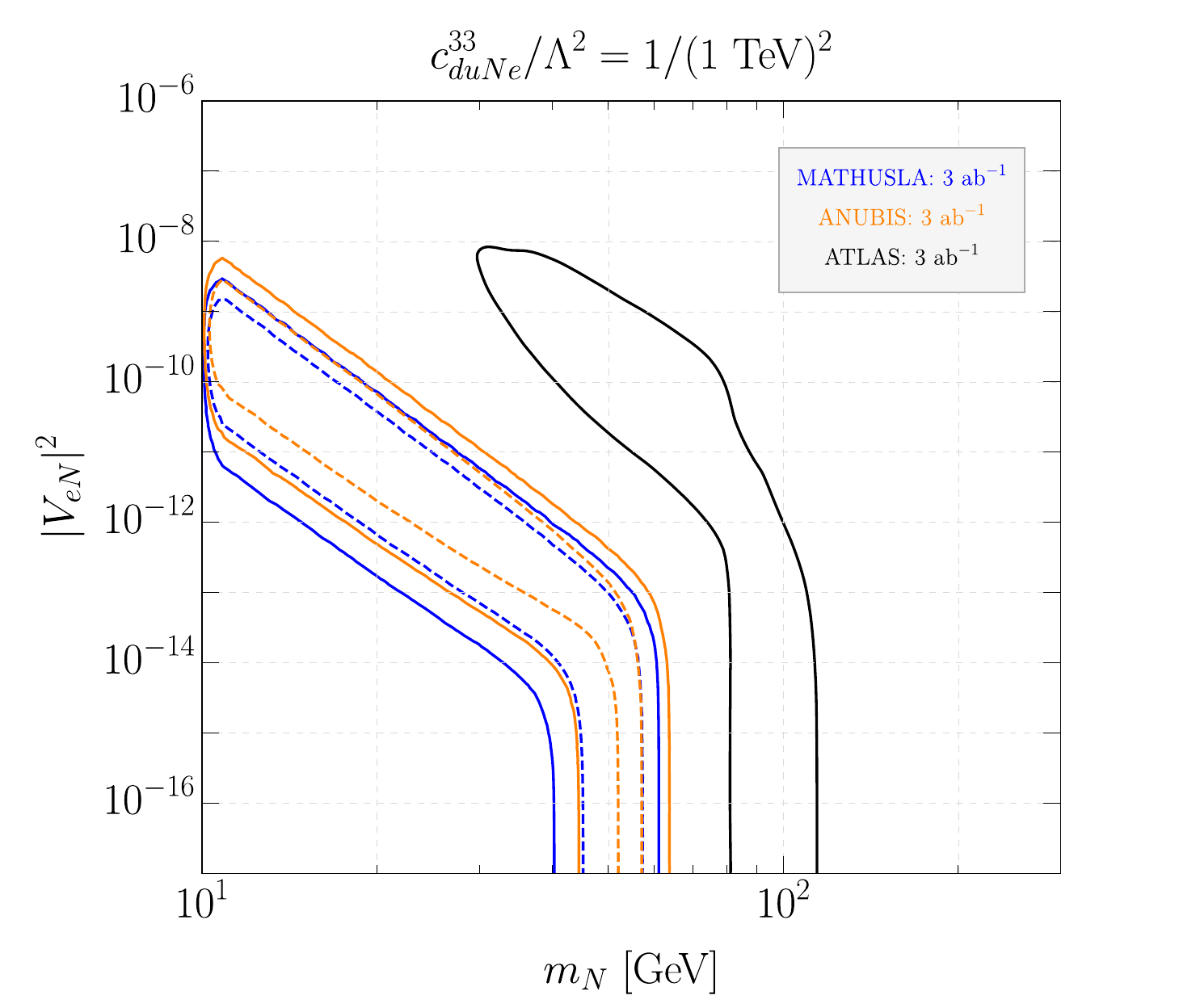} 
    \caption{Sensitivities %y limits by 
    of ATLAS and the far detectors in the plane $|V_{eN}|^2$ vs. $m_N$ to scenarios~1 (top left), 2 (top right), 3 (bottom left), and 4 (bottom right). Solid  lines correspond to 3 signal events. Dashed lines are for 30 events in all the plots, except for the lower-right panel, where they correspond to 10 events.}
    \label{fig:sensitivity_mix}
\end{figure}

In figure~\ref{fig:sensitivity_mix}, we display the sensitivity
contours of ATLAS and the far detectors in the $|V_{eN}|^2$ vs.~$m_N$
plane. The upper-left panel corresponds to scenario~1, where the
pair-$N_R$ operator structure $\mathcal{O}_{uN}^{13}$ is switched on,
assuming $c_{uN}^{13}=1$ and $\Lambda = 1$~TeV. Solid lines correspond
to 3 signal events, while dashed lines, which are present for
MATHUSLA, ANUBIS, and ATLAS, are for 30 signal
events.\footnote{Note that the smaller version of MATHUSLA, recently discussed in Ref.~\cite{mathusla_new_design} is roughly a factor 10 smaller in volume than the original MATHUSLA design~\cite{MATHUSLA:2020uve}, used in our simulation. The 30-event lines thus correspond roughly to the 3-event curves of the smaller design.}
Since the production process is controlled by the operator coefficient and decoupled from decay, these experiments can probe very small values of active-sterile-neutrino mixing and large HNL masses.
Among the far detectors, MATHUSLA and ANUBIS exhibit the strongest sensitivities, reaching HNL masses in excess of 1 TeV.
However, the EFT validity requires $m_N$ to be significantly below the new-physics scale, and we limit the plot range accordingly
for our choice of $\Lambda = 1$~TeV.
CODEX-b and MAPP2 cover smaller regions of the parameter space, but still can reach values of $|V_{eN}|^2$ as low as $10^{-20}$ and $10^{-19}$, respectively, and masses up to 800~GeV (CODEX-b) and 500~GeV (MAPP2).
Compared to the far detectors, ATLAS can probe larger values of mixing for the same HNL mass range, because of its better sensitivity to LLPs with shorter lifetimes.
The nearly $4\pi$ angular coverage further enhances its capabilities.

The sensitivities to single-$N_R$ operator scenarios are shown in the
other panels of figure~\ref{fig:sensitivity_mix}.  In the upper-right
panel, we switch on the operator $\mathcal{O}_{QuNL}^{13}$
(scenario~2), and in the lower-left and lower-right panels, we
consider the operators $\mathcal{O}_{duNe}^{13}$ (scenario 3) and
$\mathcal{O}_{duNe}^{33}$ (scenario 4), respectively.  For these
plots, we assume the respective operator coefficient
$c_\mathcal{O}^{ij} =1$ and $\Lambda = 1$~TeV. The solid and dashed
lines follow the same conventions as in scenario~1, except for the
lower-right panel, where dashed lines correspond to 10 signal events
as a result of the smaller HNL production cross-section in scenario 4 (see figure~\ref{fig:Xsecs}). In fact, there is no dashed line for ATLAS since 10 signal events are not reached in the parameter space under consideration.

In all the three single-$N_R$ operator scenarios, the sensitivity curves exhibit a funnel-like feature in a certain mass region, \textit{i.e.}~a region where the sensitivity curves become independent of the mixing parameter. 
More specifically, in this mass region, the chosen value of the operator coefficient alone leads to more than 3 (30) signal events inside the solid (dashed) contour.
In general, this funnel-like behavior appears when the operator starts to dominate the HNL decay, generating sufficient number of signal events even when the mixing parameter is negligible. 
Note, however, this does not imply sensitivity to infinitely small values of mixing: 
if the operator did not contribute to the HNL decay, the sensitivity regions in the $|V_{eN}|^2$ vs.~$m_N$ plane would be bounded from below.
The upper mass reach in these scenarios is limited by the size of the
detectors~---~for larger masses, the HNLs decay too promptly to leave
an imprint in the far detectors or produce a displaced vertex in
ATLAS.

There are qualitative differences among these scenarios.
In scenario~2 with the $\mathcal{O}_{QuNL}$ operator (upper-right panel), the sensitivity in mass is limited to $m_N \lesssim 40$~GeV for the far detectors because of the larger decay width with this operator in this mass range.
In contrast, for scenarios~3 and 4 (bottom panels), the far detectors can probe larger masses, with sensitivities up to $m_N \lesssim 80$~GeV.
On the other hand, ATLAS is sensitive to $m_N \lesssim 120$~GeV in all three scenarios.
Moreover, only the experiments with an expected integrated luminosity of $\mathcal{L} =
3$~ab$^{-1}$ (MATHUSLA, ANUBIS, and ATLAS) exhibit sensitivity to
scenarios 3 and 4.  This is because the operator $\mathcal{O}_{duNe}$
generates smaller production cross-sections compared to
$\mathcal{O}_{uN}$ and $\mathcal{O}_{QuNL}$, requiring higher
luminosities for detection.

\begin{figure}[t]
    \centering
        \includegraphics[width=0.49\linewidth]{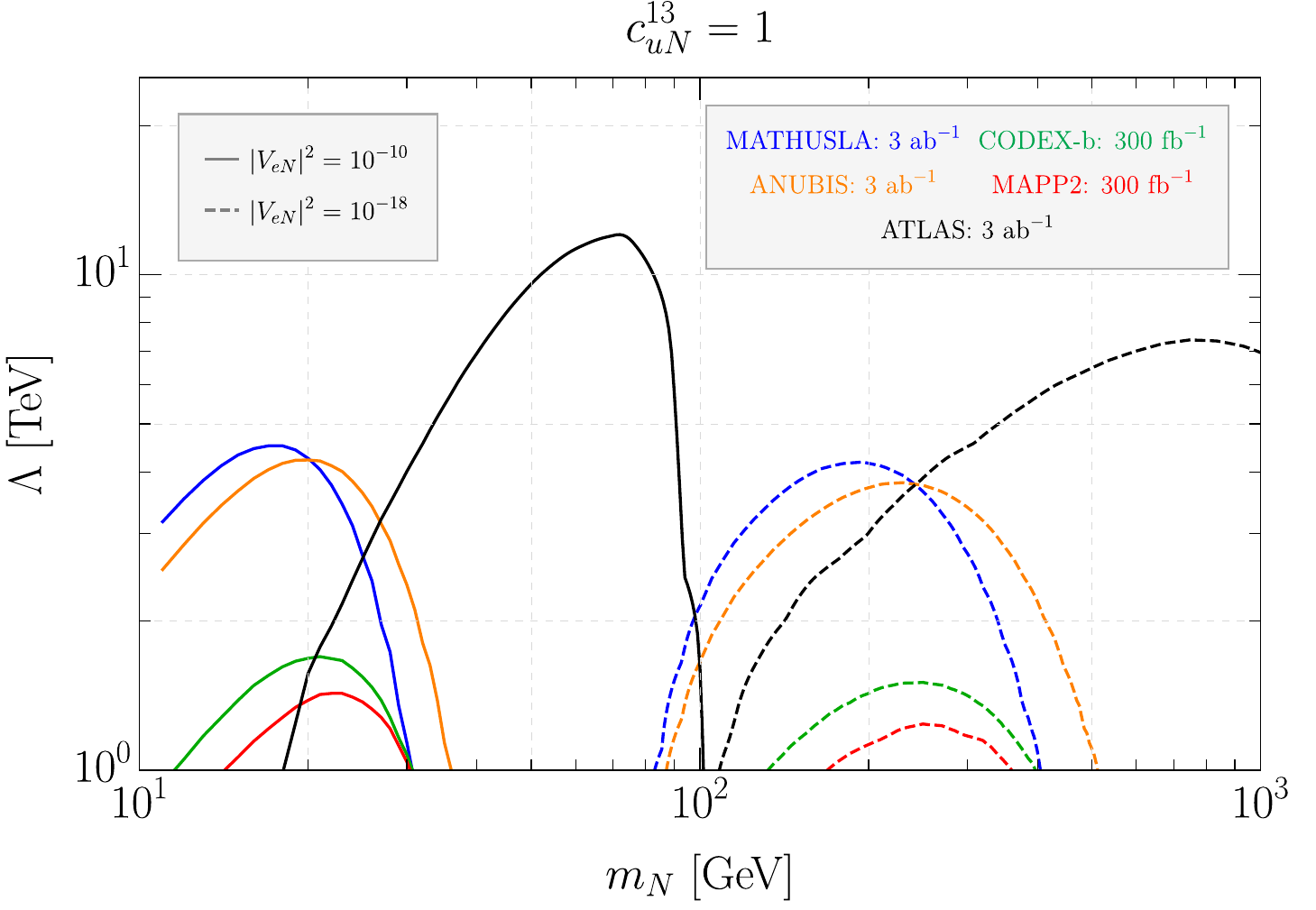} 
        \includegraphics[width=0.49\linewidth]{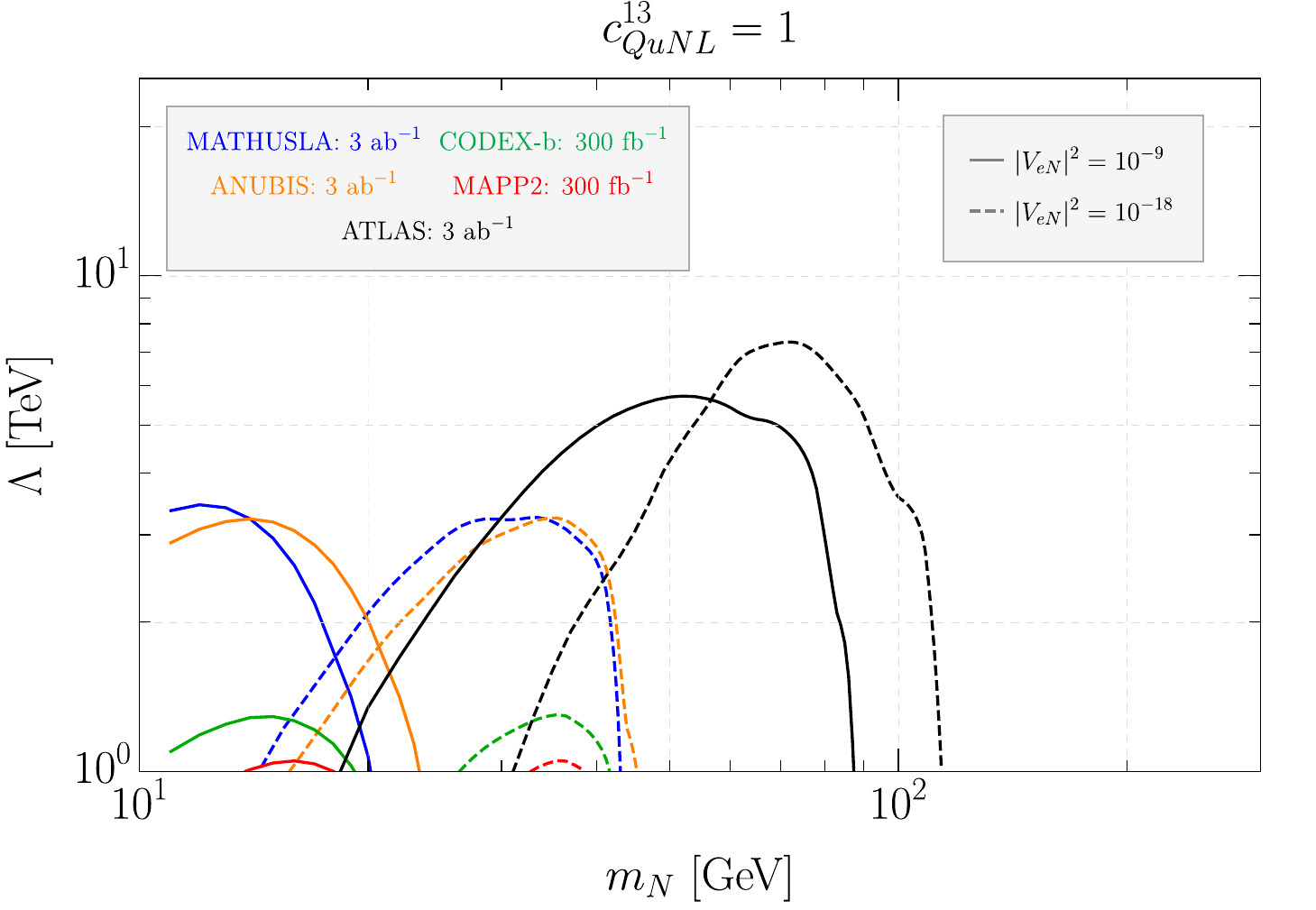} 
     \vspace{0.1cm}
         
        \includegraphics[width=0.49\linewidth]{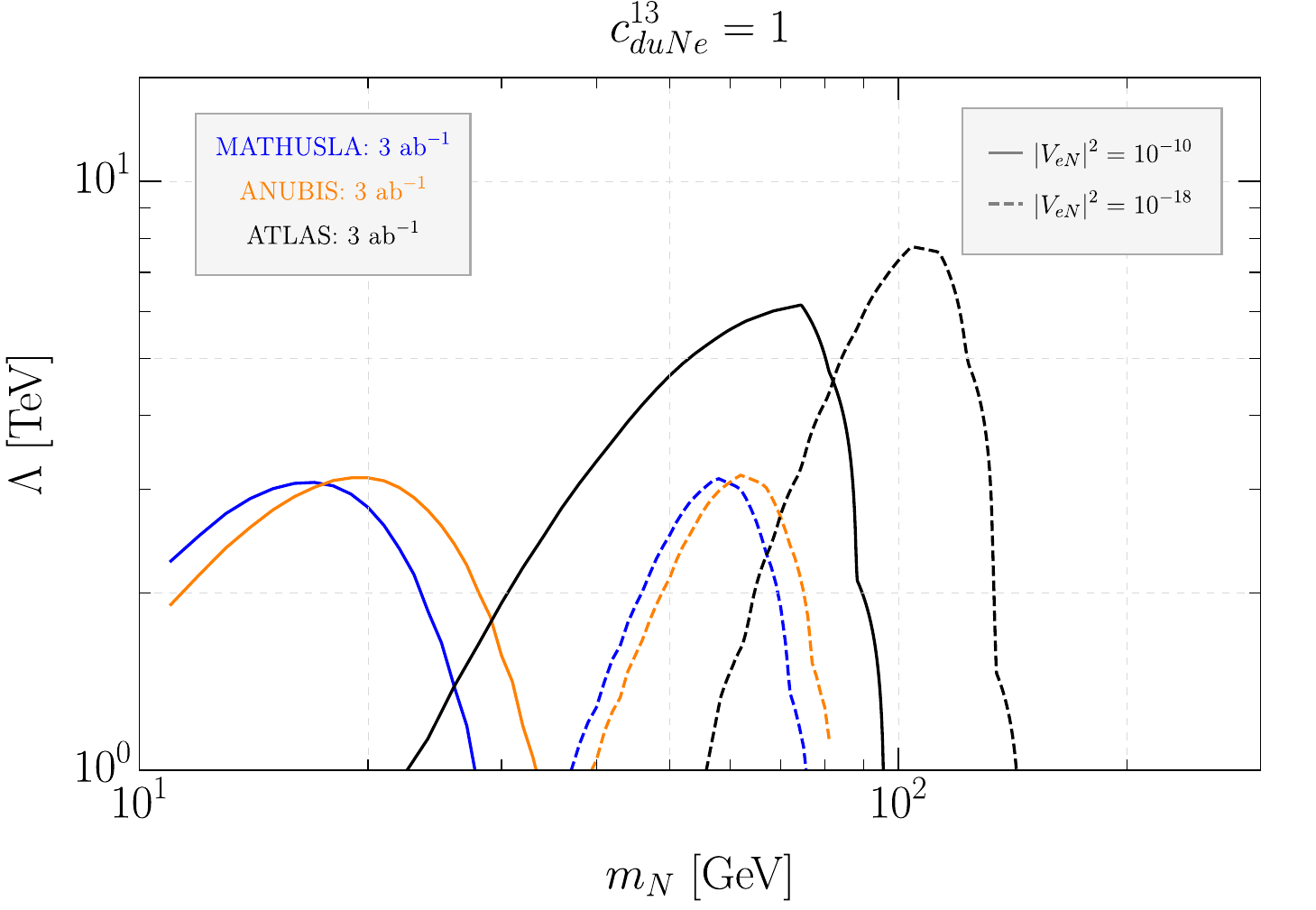} 
    \hfill
        \includegraphics[width=0.49\linewidth]{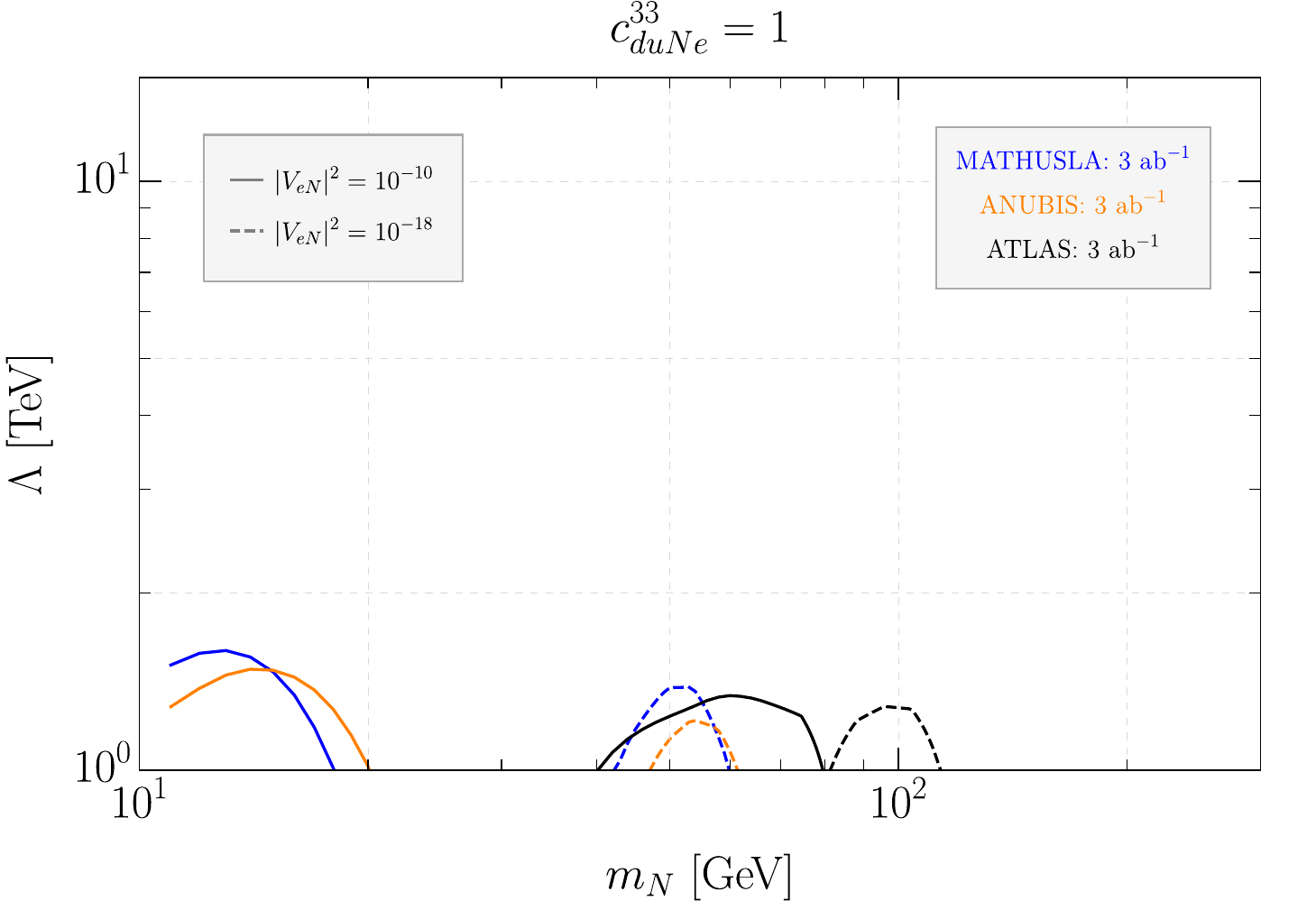} 
    \caption{Sensitivities %limits 
    of ATLAS and the far detectors in the plane $\Lambda$ vs.~$m_N$ to scenarios~1 (top left), 2 (top right), 3 (bottom left), and 4 (bottom right). Solid (dashed) lines correspond to 
    $|V_{eN}|^2 = 10^{-10}$ ($10^{-18}$), except for scenario 2, where solid lines are for $|V_{eN}|^2 = 10^{-9}$ .}
\label{fig:sensitivity_lambda}
\end{figure}

We also explore the sensitivity limits in the $\Lambda$ vs.~$m_N$ plane for fixed values of the active-sterile-neutrino mixing. 
They are displayed in the four panels of figure~\ref{fig:sensitivity_lambda}, corresponding to the operators $\mathcal{O}_{uN}^{13}$ (top left), 
$\mathcal{O}_{QuNL}^{13}$ (top right), 
$\mathcal{O}_{duNe}^{13}$ (bottom left), 
and $\mathcal{O}_{duNe}^{33}$ (bottom right). 
%The contours are shown 
We show the reach in $\Lambda$ for two mixing values: $|V_{eN}|^2 = 10^{-10}$ (solid lines)\footnote{For scenario 2, we have fixed instead $|V_{eN}|^2 = 10^{-9}$ to guarantee that the operator contribution to the HNL decay is negligible compared to the mixing one; see the bottom panel in figure~\ref{fig:widths}.} and $|V_{eN}|^2 = 10^{-18}$ (dashed lines).
The chosen values of the mixing parameter allow us to neglect 
either the operator or the mixing contribution to the HNL decay, 
such that we have a simple scaling of the number of signal events with $\Lambda$.
In all four scenarios considered, we observe that 
(i) the probed HNL masses depend strongly on the assumed value of $|V_{eN}|^2$, 
and 
(ii) for the same value of $|V_{eN}|^2$, ATLAS and the far detectors 
cover complementary mass ranges.
The accessible parameter space varies significantly between scenarios.
Scenario 1 exhibits the highest sensitivities in both $m_N$ and $\Lambda$. 
In particular, for $|V_{eN}|^2 = 10^{-10}$, ATLAS can probe the new-physics scale $\Lambda$ in excess of 10~TeV for $40~\text{GeV} \lesssim m_N \lesssim 80$~GeV. 
MATHUSLA (ANUBIS) reaches $\Lambda \approx 4.5$ (4)~TeV for $m_N$ slightly smaller than 20~GeV, whereas CODEX-b (MAPP2) can exclude $\Lambda \lesssim 1.5$ (1.2)~TeV for $m_N$ slightly larger than 20~GeV. %\AT{Please check the numbers.} 
Assuming $|V_{eN}|^2 = 10^{-18}$, a similar reach in $\Lambda$ occurs for $m_N \gtrsim 100$~GeV.
In scenario~2, ATLAS can reach $\Lambda \approx 7$~TeV (more than 8~TeV) for $m_N$ around 50 (70)~GeV if $|V_{eN}|^2 = 10^{-9}~(10^{-18})$.
MATHUSLA and ANUBIS can be sensitive to the scales as large as 3~TeV, 
whereas the sensitivities of CODEX-b and MAPP2 extend to only slightly above $\Lambda = 1$~TeV.
Scenario 3 exhibits similar sensitivities, with ATLAS being capable of probing $\Lambda$ values up to 7--8 TeV, and %the far detectors
MATHUSLA and ANUBIS being sensitive to new-physics scales of up to 3~TeV. However, the far detectors in this scenario can probe larger masses 
than those in scenario 2, for the same values of $|V_{eN}|^2$. In contrast, for scenario 4, the sensitivities are generally much weaker. Its limited reach is a result of the small production cross-sections,  which result in insufficient numbers of HNLs even for moderate values of $\Lambda$. In fact, CODEX-b and MAPP2 are not sensitive at all to scenarios~3 and 4 for the same reason, while ATLAS can only probe $\Lambda \lesssim 1.5$~TeV in a relatively small $m_N$ range in the latter scenario.

All plots in this section were calculated with the assumption that
the HNL is a Dirac particle. Let us close this section with a brief
discussion on how our results would change, if we consider Majorana
HNLs instead. There are two effects. First of all, in case of
pair-$N_R$ operators, as demonstrated in \cite{Cottin:2021lzz}, cross
sections for Majorana and Dirac HNLs are the same, except for large
values of $m_N$, say $m_N \sim {\rm(few)}$(100 GeV). At the largest
masses, for Majorana HNLs cross sections are suppressed relative to the Dirac case. Thus, the mass reach is reduced for ${\cal O}_{uN}$ relative to the values shown in figure~\ref{fig:sensitivity_mix}.
Note that figure~\ref{fig:sensitivity_mix} cuts $m_N$ at 1 TeV.  The
nature of the HNL also affects its decay width. Majorana HNLs have a
twice larger decay width, compared to Dirac HNLs at the same value
of $|V_{eN}|^2$. Thus, roughly one has to divide $|V_{eN}|^2$ by a
factor of 2 to obtain the corresponding plots for Majorana
HNLS. Note that the different plots in
figure~\ref{fig:sensitivity_mix} show ranges of $|V_{eN}|^2$ which
vary between (11--17) orders of magnitude. Thus, we consider a change
of a factor of 2 not essential. Strictly speaking this is correct
only in the limit where mixing dominates the decay. For single-$N_R$
operators, at the largest values of $m_N$ the decays are dominated
by the operator contribution. A change of this width by a factor of
two for the Majorana cases will lead then to a reduction of the
largest $m_N$, to which our search is sensitive, by roughly 15\%.

%%%%%%%%%%%%%%%%%%%%%%%%%%%%%%%%%%%%%%%%%%%

%% file: subtex/05_conclusions.tex
% !TEX root = ../top_hnl.tex
\section{Conclusions}
\label{sec:conclusions}
%%%%%%%%%%%%%%%%%%%%%%%%%%%%%
%
In the framework of 
$N_R$SMEFT, we have considered a set of the four-fermion operators with top quarks and HNLs; see table~\ref{tab:topNops}. 
Such operators may originate at low energies as a result of new heavy top-philic physics.
We have studied HNL production and decay channels 
triggered by these effective interactions at the
HL-LHC.
The resulting phenomenology depends strongly on (i)~the HNL mass, $m_N$, (ii)~the type and strength of the effective interaction encoded in the ratio $c_\mathcal{O}^{ij}/\Lambda^2$ of the Wilson coefficient and the new-physics scale $\Lambda$, as well as (iii)~the value of active-sterile-neutrino mixing $V_{e N}$.

In the present study, we have focused on long-lived, massive HNLs, that can leave a DV signature 
either in a local LHC detector, \textit{e.g.}~ATLAS or CMS, or in a planned far detector, \textit{e.g.}~MATHUSLA or ANUBIS. 
Depending on the type and the quark-flavor structure of the effective operator, HNLs can be produced at the LHC either directly in $pp$ collisions or in rare top quark decays, \textit{cf}.~figures~\ref{fig:single-top_pair-N} and \ref{fig:single-top_single-N}, and table~\ref{tab:Nproddec}. 
In the case of pair-$N_R$ operators, HNL production and decay are decoupled, since the latter can only occur via active-sterile-neutrino mixing.
For simplicity, we have assumed only one kinematically accessible HNL.
On the other hand, single-$N_R$ operators contribute to both HNL production and decay.
We have computed the corresponding production cross-sections and decay rates; see figures~\ref{fig:Xsecs} and \ref{fig:widths}, showing, in particular, how the latter compare to the decay rates induced by the active-sterile-neutrino mixing parameter $V_{e N}$.

We have further identified four benchmark scenarios possessing different phenomenology, each corresponding to a certain operator structure: $\mathcal{O}_{uN}^{13}$ in the pair-$N_R$ category, and $\mathcal{O}_{QuNL}^{13}$, $\mathcal{O}_{duNe}^{13}$, and $\mathcal{O}_{duNe}^{33}$ in the single-$N_R$ category.
We have studied in detail the prospects for probing each of these four scenarios at the HL-LHC. 
For the ATLAS main detector we have proposed a search strategy based on published ATLAS searches for LLPs, and for the LHC far detectors we simply count the visible-decay numbers of the HNLs inside their fiducial volumes.
Performing  state-of-the-art numerical simulations, 
we have shown that in the case of $\mathcal{O}_{uN}^{13}$, 
assuming $c_{uN}^{13}/\Lambda^2 = 1$~TeV$^{-2}$,
MATHUSLA (ANUBIS) can probe $|V_{eN}|^2 \approx 10^{-20}$ (and even smaller) for $m_N \gtrsim 120$ (150)~GeV,%
\footnote{Here we quote the numbers obtained under the assumption of zero background.
In figure~\ref{fig:sensitivity_mix}, we have also provided the contours corresponding 
to a larger number of signal events to give an idea for more realistic experimental conditions.} 
whereas ATLAS will be sensitive to such tiny values of the mixing for larger HNL masses, \textit{cf}.~figure~\ref{fig:sensitivity_mix}. 
For a fixed value of $|V_{eN}|^2 = 10^{-10}$, MATHUSLA (ANUBIS) will be able to exclude the new-physics scale $\Lambda \lesssim 4.5$ (4)~TeV for $m_N \approx 20$~GeV, 
whereas ATLAS will be sensitive to the scales in excess of 10 TeV for 
$40~\text{GeV} \lesssim m_N \lesssim 80$~GeV, 
as can be inferred from figure~\ref{fig:sensitivity_lambda}. 
A weaker reach in $|V_{eN}|^2$ and $\Lambda$ is observed for the single-$N_R$ scenarios. The key difference with respect to a pair-$N_R$ interaction 
is that at a certain value of $m_N$, 
a single-$N_R$ operator starts to dominate the HNL decay (with respect to the contributions from the active-sterile-neutrino mixing)
resulting in the funnel-like feature on the corresponding plots in figure~\ref{fig:sensitivity_mix}. 
The reach in $\Lambda$ depends significantly on the operator structure. 
For example, in the case of $\mathcal{O}_{duNe}^{13}$ ($\mathcal{O}_{duNe}^{33}$), ATLAS can probe the new-physics scale 
as large as 7~(1.5)~TeV for $m_N$ around 50--60~GeV.

In conclusion, the proposed DV searches at ATLAS and future far detectors at the HL-LHC, including MATHUSLA and ANUBIS, provide a very sensitive tool for probing so far unconstrained HNL-related top-philic new physics that may be hiding at a few TeV scale.